\documentclass[aps,prb,twocolumn,superscriptaddress, show pacs]{revtex4-1}
\usepackage{bm, amsmath}
\usepackage{graphicx}
\usepackage{epstopdf}
\usepackage{color}

\begin{document}

\title{
Chiral skyrmions in cubic helimagnet films: the role of uniaxial anisotropy
}

\author{M. N. Wilson}
\affiliation{Department of Physics and Atmospheric Science, Dalhousie University, Halifax, Nova Scotia, Canada B3H 3J5}

\author{A.~B.~Butenko} 
\affiliation{IFW Dresden, Postfach 270016, D-01171 Dresden, Germany}
\affiliation{Institute of Applied Physics, University of Hamburg, D-20355 Hamburg, Germany}

\author{A.~N.~Bogdanov}
\affiliation{IFW Dresden, Postfach 270016, D-01171 Dresden, Germany}

\author{T.~L.~Monchesky}
\affiliation{Department of Physics and Atmospheric Science, Dalhousie University, Halifax, Nova Scotia, Canada B3H 3J5}

\email[]{theodore.monchesky@dal.ca}
\thanks{}

\date{\today}

\begin{abstract}

This paper reports on magnetometry and magnetoresistance measurements of MnSi epilayers performed in out-of-plane magnetic fields.  We present a theoretical analysis of the chiral modulations that arise in confined cubic helimagnets where the uniaxial anisotropy axis and magnetic field are both out-of-plane.   In contrast to in-plane field measurements (Wilson \emph{et al.}, Phys. Rev. B \textbf{86}, 144420 (2012)),
the hard-axis uniaxial anisotropy in MnSi/Si(111) increases the energy of (111)-oriented skyrmions and in-plane helicoids relative to the cone phase, and makes the cone phase the only stable magnetic texture below the saturation field.  While induced uniaxial anisotropy is important in stabilizing skyrmion lattices and helicoids  in other confined cubic helimagnets, the particular anisotropy in MnSi/Si(111) entirely suppresses these states in an out-of-plane magnetic field. However,
it is predicted that isolated skyrmions with enlarged sizes exist in MnSi/Si(111) epilayers in a broad range of out-of-plane magnetic fields.
These results reveal the importance of the symmetry of the anisotropies in bulk and confined cubic helimagnets in the formation of chiral modulations and they provide additional evidence of the physical nature of the $A$-phase states in other B20-compounds.
\end{abstract}


\maketitle

\section{Introduction}

Broken inversion symmetry in magnetic crystals creates both of one-dimensional (1D) helical modulations,\cite{Dzyaloshinskii:1964jetp} and two-dimensional (2D) localized structures (\textit{chiral skyrmions}).\cite{Bogdanov:1989jetp,Bogdanov:1994jmmm}
These textures are due to \textit{Dzyaloshinskii-Moriya} (DM) interactions imposed by the chirality of the underlying crystal structure.\cite{Dzyaloshinskii:1958jpcs, Moriya:1960pr}
Similar interactions in other condensed matter systems that lack
inversion symmetry (such as multiferroics,\cite{Seki:2012sci} ferroelectrics, chiral liquid crystals\cite{Fukuda:2011nc}) can also stabilize skyrmionic states.\cite{Bogdanov:1995jetpl}
Importantly, multi-dimensional solitons are unstable in most achiral nonlinear systems and collapse spontaneously into point or linear singularities. \cite{Derrick:1964jmp}
This fact attaches special importance to chiral condensed matter systems as a particular class of
materials where skyrmion states can exist. 

Among noncentrosymmetric magnetic compounds, easy-axis ferromagnets with $n$mm ($C_{nv}$) and $\bar{4}2m$ ($D_{2d}$) symmetries can be considered as the most suitable crystals to observe chiral skyrmions. In these compounds, condensed 2D chiral skyrmion textures (\textit{skyrmion lattices}) can exist as thermodynamically stable states in a broad range of applied magnetic fields and temperatures \cite{Bogdanov:1989jetp, Bogdanov:1994jmmm}. 
In other classes of noncentrosymmetric magnets, skyrmion lattices compete with one-dimensional modulations (\textit{helicies}) and arise only for certain ranges of the material parameters.  
 A number of recent investigations indicate the possible existence of chiral skyrmions in noncentrosymmetric uniaxial ferromagnets.\cite{Ghimire:2013prb}
In cubic helimagnets, the situation is even more difficult for skyrmion formation: one-dimensional single-harmonic modulations (\textit{cone} phases) correspond to the global energy minimum in practically the entire region where chiral modulations exist, while skyrmion lattices and helicoids can exist only as metastable states.\cite{Butenko:2010prb, Wilhelm:2012jpcm} 
Skyrmionic states and other multidimensional modulated textures are reported to exist only in close vicinity to the Curie temperatures ($T_{C}$) of bulk cubic helimagnets as so called \textit{precursor states}. \cite{Kusaka:1976ssc, Kadowaki:1982jpsj, Lebech:1995jmmm, Grigoriev:2006prb2, Stishov:2008jetp,  Muhlbauer:2009sci, Pappas:2009prl,Wilhelm:2011prl, Wilhelm:2012jpcm, Seki:2012prb1}

Beyond the precursor region, condensed skyrmion phases and other thermodynamically stable nontrivial modulations are expected to exist only in cubic helimagnets where additional stabilizing effects are present. Theoretical analysis and experimental observations show that surface/interface induced uniaxial distortions \cite{Butenko:2010prb,Karhu:2012prb} and finite size effects \cite{Rybakov:2013prb, Wilson:2013prb} effectively suppress unwanted cone states and stabilize helicoids and skyrmion lattices in confined cubic helimagnets. 
Recently, the challenge of creating and observing such textures was overcome by the fabrication of free-standing nano-layers of cubic helimagnets\cite{Yu:2010nat} and the synthesis of epitaxial thin films of these materials on Si(111) substrates.\cite{Karhu:2010prb, Karhu:2011prb, Wilson:2012prb, Huang:2012prl, Porter:2012prb, Wilson:2013prb, Engelke:2013jpcm}
Despite numerous indirect indications of skyrmionic states in different nonlinear systems \cite{Brown:2010, Volovik:1992} confined cubic helimagnets still remain the only class of materials where  skyrmionic states can be induced, observed, and manipulated in a broad range of the thermodynamical parameters.
\cite{Yu:2010nat, Yu:2011nm, Wilson:2012prb, Huang:2012prl,Tonomura:2012nl}
Investigations of chiral skyrmions in cubic helimagnet nano-layers have gained importance after the discovery of similar skyrmionic states stabilized by surface/interface DM interactions in nano-layers of common magnetic metals \cite{Heinze:2011np, Romming:2013sci} and perspectives of their applications in data storage technologies.\cite{Kiselev:2011jpd,Romming:2013sci, Fert:2013nn}

  Anisotropy plays a decisive role in the structure of skyrmions in epitaxial films of cubic helimagnets.  Due to the lattice mismatch between the B20 crystal and the Si(111) substrate, strain induces a uniaxial magnetic anisotropy through magnetoelastic coupling.  This uniaxial anisotropy can lead to two kinds of regular skyrmions. The (111)-easy-plane uniaxial anisotropy in MnSi/Si(111) stabilizes skyrmions with their cores lying along the in-plane direction,\cite{Wilson:2012prb} whereas the (111)-easy-axis anisotropy in FeGe/Si(111) produces skyrmions with their cores aligned along the [111] direction.\cite{Huang:2012prl}

In addition, specific effects imposed by a confined geometry of nano-layers also contribute to the stability of complex magnetic textures over a broad range of thermodynamic parameters.\cite{Rybakov:2013prb, Wilson:2013prb}
We address these finite-size effects in a separate paper.\cite{Rybakov:unpub} 
In this work we concentrate on effects imposed by induced uniaxial distortions. 

In our previous polarized neutron reflectometry (PNR) and magnetometry study, we showed that the ground state of MnSi thin films in a thickness range 7 nm $\leq d \leq$ 40 nm is helimagnetic with a propagation vector oriented along the out-of-plane [111] direction.  
Measurements with both techniques yield a helical wavelength of $L_D = 13.9$~nm. \cite{Karhu:2011prb}
Following the introduction of a new class of magnetic materials in the form of epilayers of cubic helimagnets in Refs.~\onlinecite{Karhu:2010prb,Karhu:2011prb}, we conducted detailed investigations of the magnetic states in MnSi/Si(111) films for in-plane magnetic fields.\cite{Karhu:2012prb, Wilson:2012prb, Wilson:2013prb}

This paper investigates the magnetic properties of MnSi/Si(111) nano-layers in out-of-plane magnetic fields.
We extend earlier calculations\cite{Bogdanov:1994jmmm,Butenko:2010prb} to include solutions for basic chiral modulations in cubic helimagnets with hard-axis uniaxial distortions and construct the magnetic phase diagrams of the solutions (Section \ref{sec:theory}).

The measurements in an out-of-plane magnetic  field  confirm the absence of thermodynamically stable (111)-skyrmions lattices and in-plane helicoidal phases (Section~\ref{sec:exp}).  This conclusion is supported by theoretical calculations, which demonstrate that the cone phase is the only thermodynamically stable phase below the saturation field over the entire magnetic phase diagram.  We explain the difference in the behavior of epilayers and bulk crystals in Section~\ref{sec:dis}.  In Section~\ref{sec:revisit} we compare the magnetization processes observed in bulk to those in confined cubic helimagnets and update the $T-H$ phase diagram.

Finally, we overview the existing observations in confined chiral systems within
the framework of our results (Section V).

\section{Chiral modulations in cubic helimagnets with uniaxial distortions}
\label{sec:theory} 

\begin{figure}[!hb]
\centering
\includegraphics[width=87 mm]{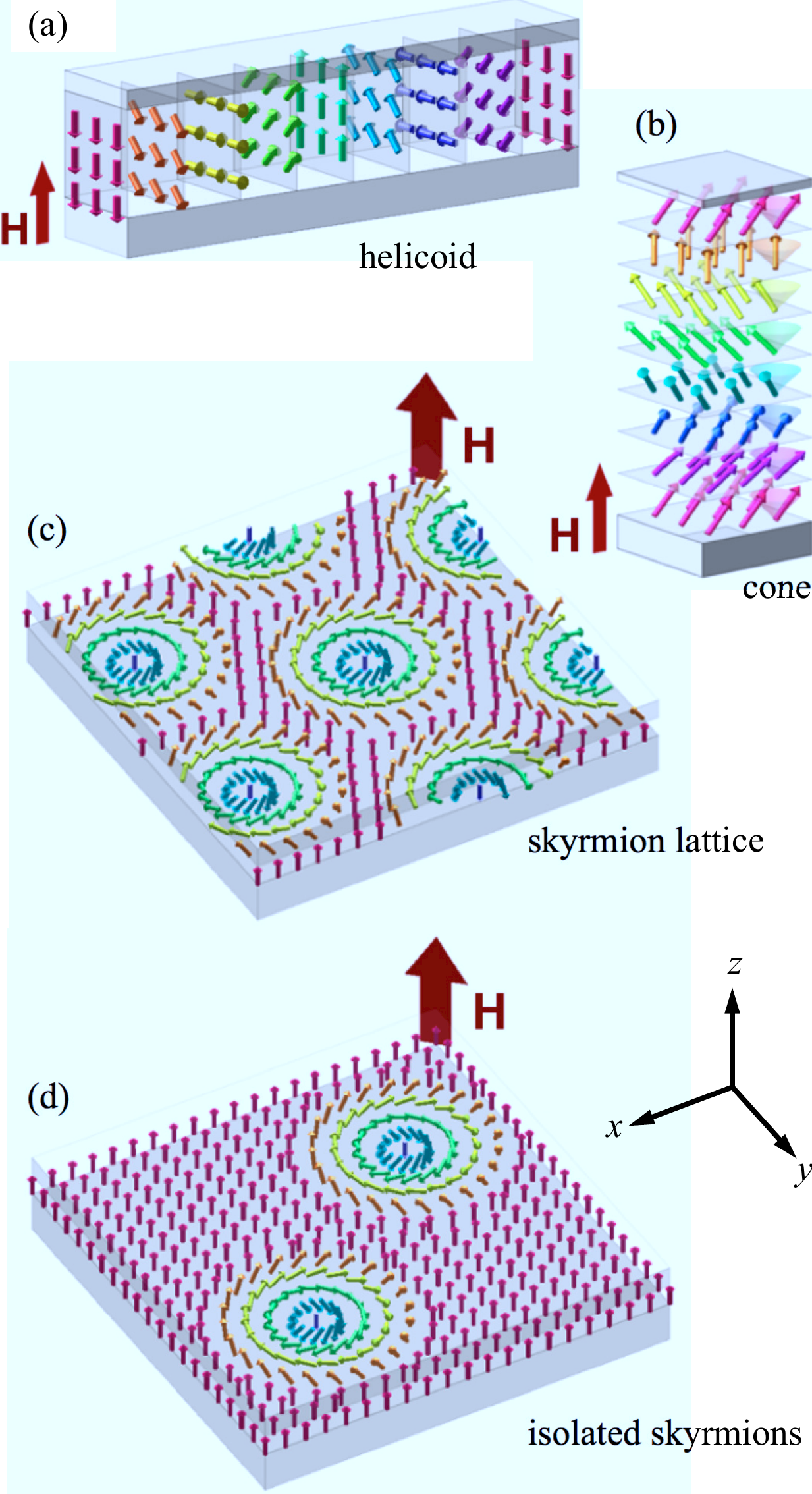}
\caption{ . (a,b) one-dimensional and (c,d) two-dimensional chiral modulations that can exist as either stable or metastable states.}
\label{fig:phases}
\end{figure}

Modulated states that arise in cubic helimagnets have been described within the Dzyaloshinskii theory of chiral helimagnets \cite{Dzyaloshinskii:1964jetp} in Refs.~\onlinecite{Bak:1980jpc, Nakanishi:1980ssc}.
The energy functional introduced by Bak and Jensen \cite{Bak:1980jpc} became the basic model and formed a conceptual framework for magnetism of cubic helimagnets.

It is well-established that a strong uniaxial magnetic anisotropy arises in epilayers of cubic helimagnets as a result of surface/interface interactions and epitaxially induced strain.\cite{Karhu:2010prb, Karhu:2011prb, Karhu:2012prb, Huang:2012prl, Porter:2012prb}
The magnetic states in theses systems can be derived by minimization of a Bak-Jensen functional (Eq.~(\ref{density}) in Ref.~\onlinecite{Bak:1980jpc}) that includes an additional uniaxial anisotropy with constant $K$.\cite{Butenko:2010prb, Karhu:2012prb} 
In this paper we write the energy density  $w (\mathbf{M})$ for a cubic helimagnet nano-layer in a magnetic field perpendicular to the film surface ($\mathbf{H} || \mathbf{z}$) as  a sum  of three energy contributions, $w = w_0 (\mathbf{M}) + w_c (\mathbf{M}) +f(M)$:
\begin {eqnarray}
w_0 (\mathbf{M}) = A \left(\mathrm{grad}\,\mathbf{M} \right)^2
-D\,\mathbf{M}\cdot \mathrm{rot}\mathbf{M}
- H M_z - K M_z^2, \ \:\ \ 
\label{density}
\end{eqnarray}
\begin {eqnarray}
w_c (\mathbf{M}) =  \sum_{i=1}^{3}
\left[B(\partial M_i / \partial x_i )^2 +B_c M_i^4 \right], 
\label{cubic}
\end{eqnarray}
\begin {eqnarray}
f (M) = J(T-T_0)M^2 + bM^4. \ \:\ \ 
\label{f}
\end{eqnarray}
The functional $w_0 (\mathbf{M})$ describes the main magnetic interactions in terms of the exchange interaction with exchange stiffness constant $A$,  the Dzyaloshinskii-Moriya (DM) coupling with constant $D$, the Zeeman energy, and the induced uniaxial anisotropy.
The energy contribution $w_c$ includes exchange anisotropy ($B$) and 
cubic magnetocrystalline anisotropy ($B_c$) 
(the $x_i$ are the components of the spatial variable).\cite{Bak:1980jpc}  
In cubic helimagnets, exchange and magnetocrystalline anisotropies 
are much smaller than the interactions
included in $w_0$. 
The energy density $f(M)$ comprises
magnetic interactions imposed by
the variation of the magnetization
modulus $M \equiv |\mathbf{M}|$ and
is written in the spirit of the Landau theory as an
expansion of the free energy with respect to
 the order parameter $M$,
 with coefficients $J$ and $b$. \cite{Bak:1980jpc}
The characteristic temperature $T_0$ is related to 
the Curie temperature of a cubic helimagnet,
$T_C = T_0 + D^2/(4JA)$.\cite{Rossler:2006nat} 
In a broad temperature range,
the magnetization vector practically does not
change its length, and the non-uniform magnetic states
 only include a rotation of $\mathbf{M}$.  Spatial modulations of the
magnetization modulus become a sizeable effect in the \textit{precursor} region and lead to the
specific effects observed in this region in close vicinity to $T_C$  (see Ref.~\onlinecite{Wilhelm:2012jpcm}
and the bibliography therein). 

In this paper we investigate a model
that has a fixed magnetization
modulus $M$ = const, Eq.~(\ref{density}). We discuss possible distortions 
of the basic magnetic phases imposed by cubic anisotropy, 
stray-fields, and spatial variations of $M$ at the end of the paper. 

For $w_0$ with $\mathbf{H} || \mathbf{z}$ and easy-axis anisotropy ($K > 0$), the solutions for one-dimensional and two-dimensional chiral modulations include the \textit{cone} phase, the \textit{helicoid}, isolated \textit{skyrmions}, and \textit{skyrmion lattices} (Fig. \ref{fig:phases}). \cite{Dzyaloshinskii:1964jetp, Bak:1980jpc,
Bogdanov:1994jmmm, Butenko:2010prb} 
In this section we investigate Eq.~(\ref{density}) with $\mathbf{H} || \mathbf{z}$ and $K<0$, which describes chiral modulations in MnSi/Si(111) films with a hard-axis anisotropy, and construct the phase diagram of the solutions for functional (\ref{density}) (Fig. \ref{fig:phasediagram}).
\begin{figure}[h]
\centering
\includegraphics[width=78 mm]{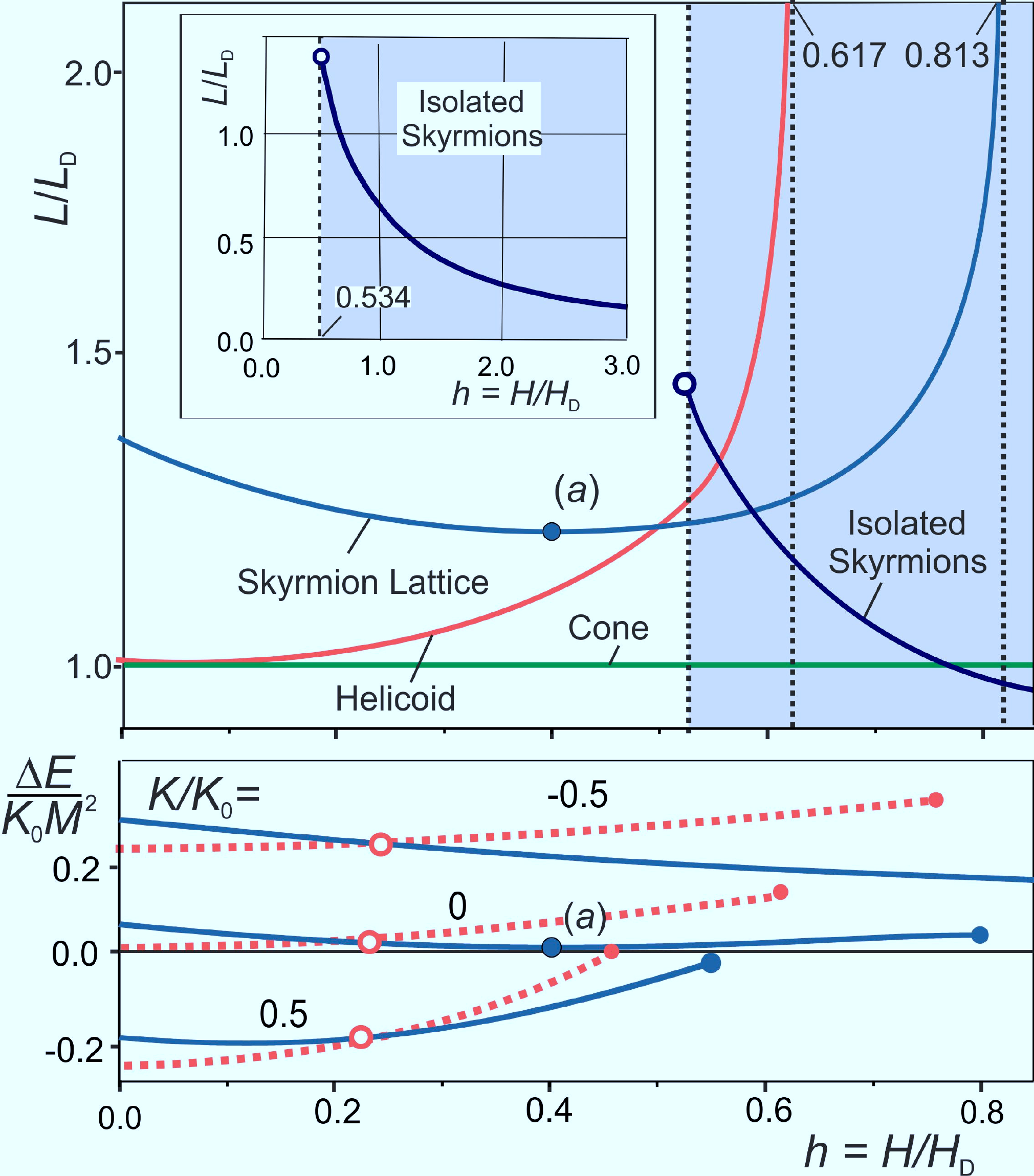}
\caption{ (color on-line) The equilibrium periods
of the modulated phases  shown in Fig. \ref{fig:phases} and isolated skyrmion sizes as a function of the applied field for $K = 0$. This is representative of the main features of chiral modulations in films with $K >0$ and $K<0$, and shows the transition of the helicoids into a set of isolated domain walls (\textit{kinks}) at $H_h = 0.617 H_D$ and the skyrmion lattice into a ``gas'' of isolated skyrmions at $H_s = 0.813 H_D$.\cite{Dzyaloshinskii:1964jetp,Bogdanov:1994jmmm}
Isolated skyrmions exist above the elliptical instability field $H_{el} = 0.534 H_D$ indicated by the grey shaded region.\cite{Bogdanov:1994pss}
Inset shows the equilibrium skyrmion sizes at high magnetic fields.
The lower panel  shows the differences between the energy densities for the helicoids (solid lines) and the skyrmion lattice (dashed lines) relative to the cone phase as functions of the applied field.
}
\label{fig:solutions2}
\end{figure}

\begin{figure}[h]
\centering
\includegraphics[width=87 mm]{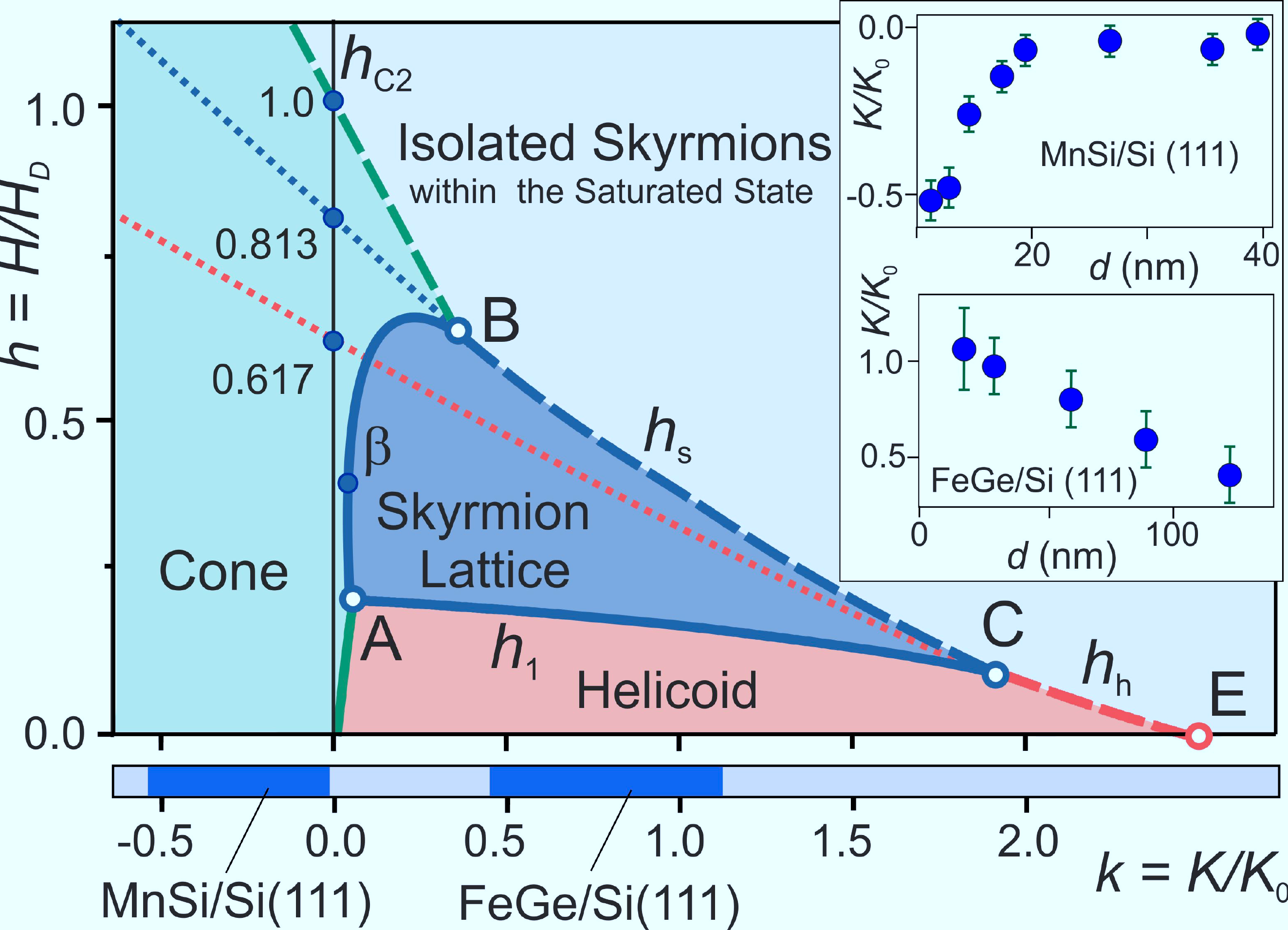}
\caption{(color on-line) The phase diagram of the equilibrium states for model (\ref{density}) with the two control parameters of model (\ref{density}), in reduced values of applied magnetic field $h = H/H_D$  ($\mathbf{H}\|\mathbf{\hat{z}}$) and uniaxial anisotropy $k = K/K_0$, as independent variables (the details are given in  Fig. \ref{fig:app1}). 
Filled areas indicate the regions of global stability for the cone (green), helicoid (red), and skyrmion lattice (blue). In the saturated state (grey area), skyrmions exist as isolated (noninteracting) objects. Thin dotted lines designate critical lines for the metastable helicoid ($h_h$) and skyrmion lattice ($h_s$); $h_1 (k)$ is the first-order transition line between the stable helicoid and skyrmion lattice. 
Inset shows the induced uniaxial anisotropy as a function of the film thickness in
hard-axis MnSi/Si(111)\cite{Karhu:2012prb} and easy-axis FeGe/Si(111) epitaxial films.\cite{Huang:2012prl}
The induced anisotropy ranges for these compounds are indicated along the $K/K_0$ axis.
}
\label{fig:phasediagram}
\end{figure}

(1) \underline{Conical helices}. By introducing spherical coordinates for the magnetization vector,
\begin {eqnarray}
\mathbf{M} = M (\sin \theta \cos \psi, \sin \theta \sin \psi, \cos \theta),
\label{sc}
\end{eqnarray}
 one can readily derive analytical solutions for the cone phase, 
\cite{Bak:1980jpc,Butenko:2010prb}

\begin {eqnarray}
\cos \theta = \frac{H}{H_{C2}},\  \psi = \frac{2\pi z}{L_D}, \ 
H_{C2} = H_D \left( 1 - \frac{K}{K_0} \right),\ \
\label{cone}
\end{eqnarray}
where
\begin {eqnarray}
L_D = 4 \pi A/|D|, \quad H_D = D^2 M/(2A).
\label{period}
\end{eqnarray}
The helical wavelength at zero field and zero anisotropy ($L_D$) and the saturation field of the cone phase ($H_D$) for $K = 0$ represent two of the characteristic material parameters of cubic helimagnets (see Table 1 in Ref.~\onlinecite{Wilhelm:2012jpcm}).

As the field increases along the propagation direction, the spins cant towards the field and produce the single-harmonic modulation described by Eqs.~(\ref{cone}) and shown in Fig.~\ref{fig:phases}(b).  The magnetic field competes with the DM-interaction, which  is represented by the effective easy-plane anisotropy $K_0 = D^2/(4A)$, and transforms the cone continuously into the saturated state ($\theta = 0$) at the critical field  $H_{C2} (K)$.	

(2) \underline{Helicoids}. 
Distorted helical modulations, known as helicoids, are shown in Fig. \ref{fig:phases} (a) for the case of an in-plane propagation vector.
The transverse distortions imposed by applied magnetic fields and/or uniaxial anisotropy are described by solutions to the well-known differential equations for the non-linear pendulum.\cite{Dzyaloshinskii:1964jetp,Wilson:2013prb}

In bulk helimagnets, a helicoid evolves continuously from a single-harmonic helix with period $L_D$ (Eq.~(\ref{period})) into a one-dimensional soliton lattice at high fields.\cite{Dzyaloshinskii:1964jetp} The lattice transforms into a set of isolated domain walls (\text{kinks}) at a critical field $H_{h} (K)$ (Figs.~\ref{fig:solutions2} and \ref{fig:phasediagram}).
This result is achieved by ignoring the weak demagnetizing field contribution.  Contrary to the case $\mathbf{H} \perp \mathbf{z}$ described in Ref.~\onlinecite{Wilson:2013prb}, these helicoids would have a continuous field dependence similar to bulk crystals. 

\begin{figure}[ht!]
\centering
\includegraphics[width= 70 mm]{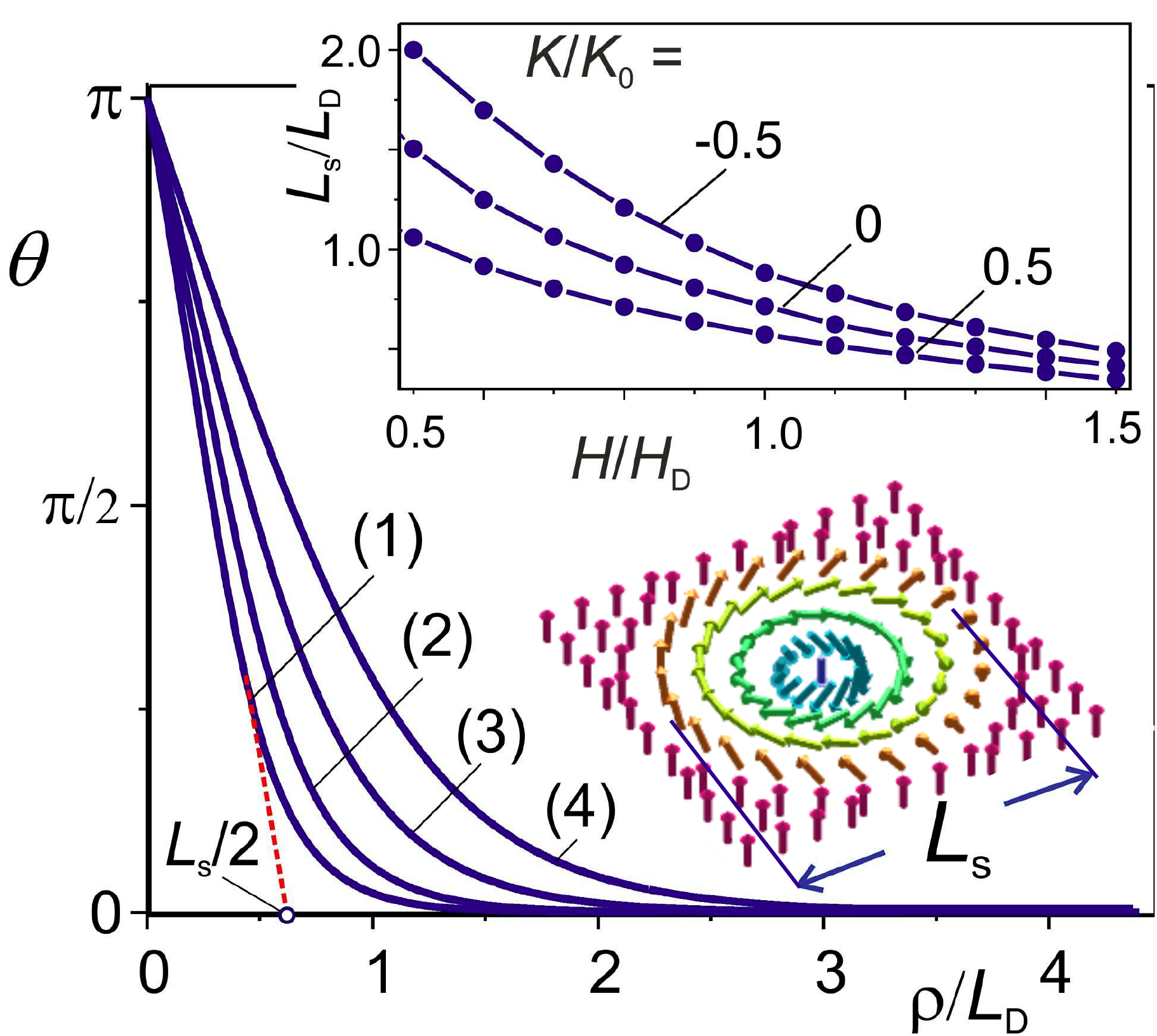}
\caption{Solutions of Eq.~(\ref{skyrmion}) for isolated skyrmions with the control parameters ($K/K_0$, $H/H_D$): (1) - ( 0.5, 1.0); (2) -(0, 1.0); (3) (-0.5, 1.0); (4) - (-0.5, 0.8). Inset shows the skyrmion sizes in films with different types of uniaxial anisotropy and different values of the applied magnetic field. }
\label{fig:solutions}
\end{figure}

(3) \underline{Isolated and embedded skyrmions}. 
For a magnetization given in spherical coordinates (Eq.~\ref{sc}), and the spatial variables in cylindrical coordinates,
$\mathbf{r} = (\rho \cos \varphi, \rho \sin \varphi, z)$, axisymmetric localized solutions (isolated skyrmions) for  Eq.~(\ref{density}) are described by $\psi = \varphi + \pi/2$ and $\theta = \theta (\rho)$, which are derived from the Euler equation,
\begin {eqnarray}
\frac{d^2 \theta}{d \rho^2} &+& \frac{1}{\rho} \frac{d \theta}{d \rho} 
-\frac{1}{\rho^2} \sin \theta \cos \theta 
+ \frac{2}{\rho} \sin^2 \theta \quad
\nonumber \\
&-& (K/K_0) \sin \theta \cos \theta - (H/H_D) \sin \theta = 0,
\label{skyrmion}
\end{eqnarray}
with boundary conditions, $\theta (0) = \pi$, $\theta (\infty) = 0 $.\cite{Bogdanov:1989jetp,Bogdanov:1994jmmm}
Typical solutions $\theta (\rho)$ for negative
$K$ are plotted in Fig. \ref{fig:solutions} together with magnetization 
profiles for isotropic ($K = 0$) and easy-axis ($ K = 0.5 K_0$) helimagnets.

Analysis shows that in a broad range of control parameters, chiral modulations are qualitatively similar in helimagnets with different signs of $K$.  For the case $K=0$, the shaded region in Fig.~\ref{fig:solutions2} shows the fields where isolated skyrmions form.  At the highest fields, the field induced saturated state is the lowest energy state, although isolated skyrmions can form inside this phase. When the field is lowered to $H_s(0) = 0.813 H_D$, metastable skyrmion lattices are able to condense.  Then, as the field reaches $H_h(0) = 0.617 H_D$, isolated domain walls condense into metastable helicoids, while skyrmion lattices and isolated skyrmions remain as metastable solutions.  Below the strip-out field $H_{el} (0) = 0.534H_D$, the isolated skyrmions become unstable and collapse into the stable helicoid phase.\cite{Bogdanov:1994pss}

Ensembles of isolated skyrmions have been observed in Fe$_{0.5}$Co$_{0.5}$Si mechanically thinned films \cite{Yu:2010nat} and FePd nano-layers \cite{Romming:2013sci} as a result of a skyrmion lattice expansion in a high magnetic field ($H > H_s$).
The equilibrium energy densities of the skyrmion lattice ($\Delta E_s = E_s - E_c$) and the helicoid ($\Delta E_h = E_h-E_c$) relative to that of the cone phase ($E_c$) are calculated from the model given by Eq.~(\ref{density}) and  are plotted in the lower panel of Fig.~\ref{fig:solutions2} as a function of the reduced magnetic field,  $h = H/H_D$, for the isotropic case ($K = 0$), as well as the easy-axis ($K = 0.5K_0$) and hard-axis ($K = -0.5K_0$) anisotropies.

The $K-H$ phase diagram in Fig. \ref{fig:phasediagram} (see also Fig.~\ref{fig:app1} in the Appendix) overviews magnetic properties of confined cubic helimagnets with different signs of uniaxial anisotropy in perpendicular magnetic fields. A corresponding phase diagram for hard-axis systems in in-plane magnetic fields has been constructed in Ref.~\onlinecite{Karhu:2012prb} and has been applied to analyze magnetic states in MnSi/Si films.

\section{ Experimental results}
\label{sec:exp}

\subsection{Sample Preparation}

\begin{figure}[!ht]
\centering
\includegraphics[width = 78 mm]{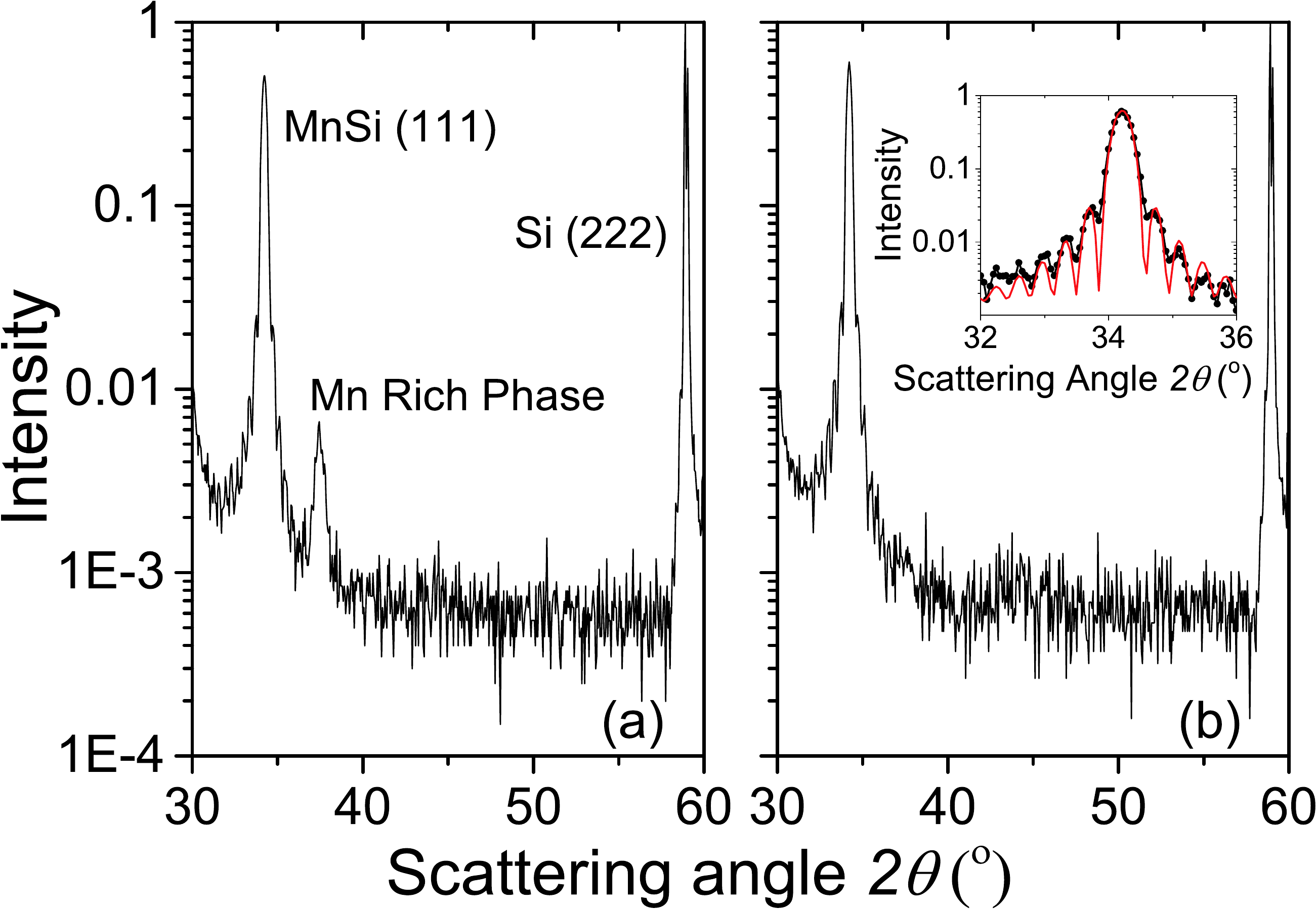}
\caption{XRD curves of the 25.4~nm sample before (a), and after (b) annealing. In both figures the intensity is normalized to the maximum height of the Si(222) substrate peak which remains unchanged through the annealing.  The inset shows a fit to the Kiessig fringes.}
\label{fig:XRD}
\end{figure}

The 25.4-nm thick MnSi thin film was grown on a Si(111) high resistivity wafer ($\rho \ge 5$k$\Omega$-cm) by co-deposition of Mn and Si, as described in Ref.~\onlinecite{Karhu:2011prb}.  This film is representative of MnSi films in a range of thicknesses from $ 12 \le d \le 40$ as our previous work has shown that the magnetic behaviour is qualitatively similar in this range.  The 25.4-nm sample was annealed \emph{ex-situ} under an Ar atmosphere for one hour at $400~^{\circ}$C to transform the residual manganese rich phase that was present in the film into MnSi. X-ray diffraction (XRD) measurements in the region $2\theta = 30^{\circ} - 60^{\circ}$ presented in Fig.~\ref{fig:XRD}(b) show no detectable impurity phase in this sample after the annealing and the Kiessig fringes in the inset demonstrate the high interfacial quality and uniformity of the film.  This annealing did not affect the magnetic states of the film other than to increase the saturation magnetization due to the increased MnSi volume. 

\begin{figure}[!t]
\includegraphics[width = 78 mm]{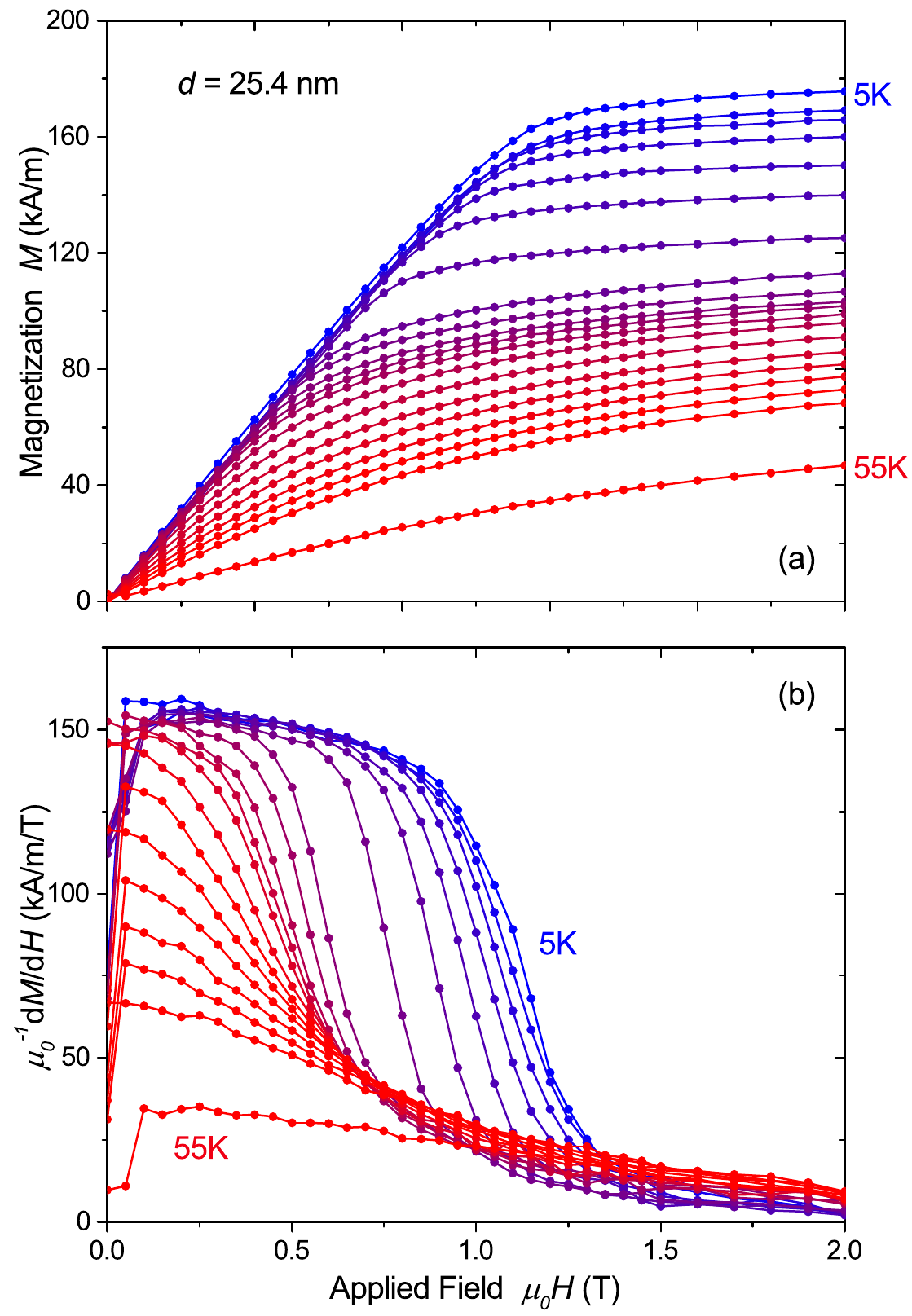}
\caption{(color on-line) (a) Magnetization curves from a $d=25.4$~nm MnSi/Si(111) film with $\mathbf{H}\| [111]$. Temperatures shown are 5, 10, 15, 20, 25, 30, 35, 40, 41, 42, 42.5, 43, 44, 45, 46, 47, 48, 49, 50 and 55~K. (b) The static susceptibility is obtained by calculating $dM/dH$ from the the data in (a). }
\label{fig:hys}
\end{figure}

As a second check of the sample quality, we determined the residual resistivity ratio (RRR) between $T = 299$~K and $T = 2$~K from magnetoresistance (MR) measurements. For these measurements, we photolithographically patterned a portion of the MnSi film into a Hall-bar using SPR220-3.0 photoresist and Ar-ion etching.  We then attached Au-wire leads onto the surface using In solder for four point resistivity measurements. The high RRR = 26.8 is further evidence of the high sample quality.    

\subsection{Magnetometry}
 
We explored the phase diagram by measuring the magnetization $M$ as a function of applied magnetic field $H$ and as a function of temperature $T$ with a Quantum Design MPMS-XL SQUID magnetometer with the applied magnetic field pointing out of plane along the MnSi [111] direction. 
The magnetic susceptibility is a common tool for mapping the phase diagram in magnetic systems.\cite{Baryakhtar:1988spu, Bogdanov:2003prb}  In bulk cubic helimagnets, peaks in the magnetic susceptibility ($dM/dH$) are signatures of the first-order magnetic phase transitions that separate the cone phase from two adjacent areas: the low-field region with multi-domain helical states  and  a small closed region near the ordering temperature, the $A$-phase pocket (Fig. \ref{fig:Hc2}(b)).\cite{Kadowaki:1982jpsj, Wilhelm:2011prl, Seki:2012prb1}  The first-order character of the transition in and out of the $A$-phase is a reflection of the differences in the topology between these phases.  In contrast, a second-order transition, identified by a minimum in $d^2M/dH^2$, exists between the conical phase and the field induced ferromagnetic state.

\begin{figure}[!t]
\includegraphics[width=87 mm]{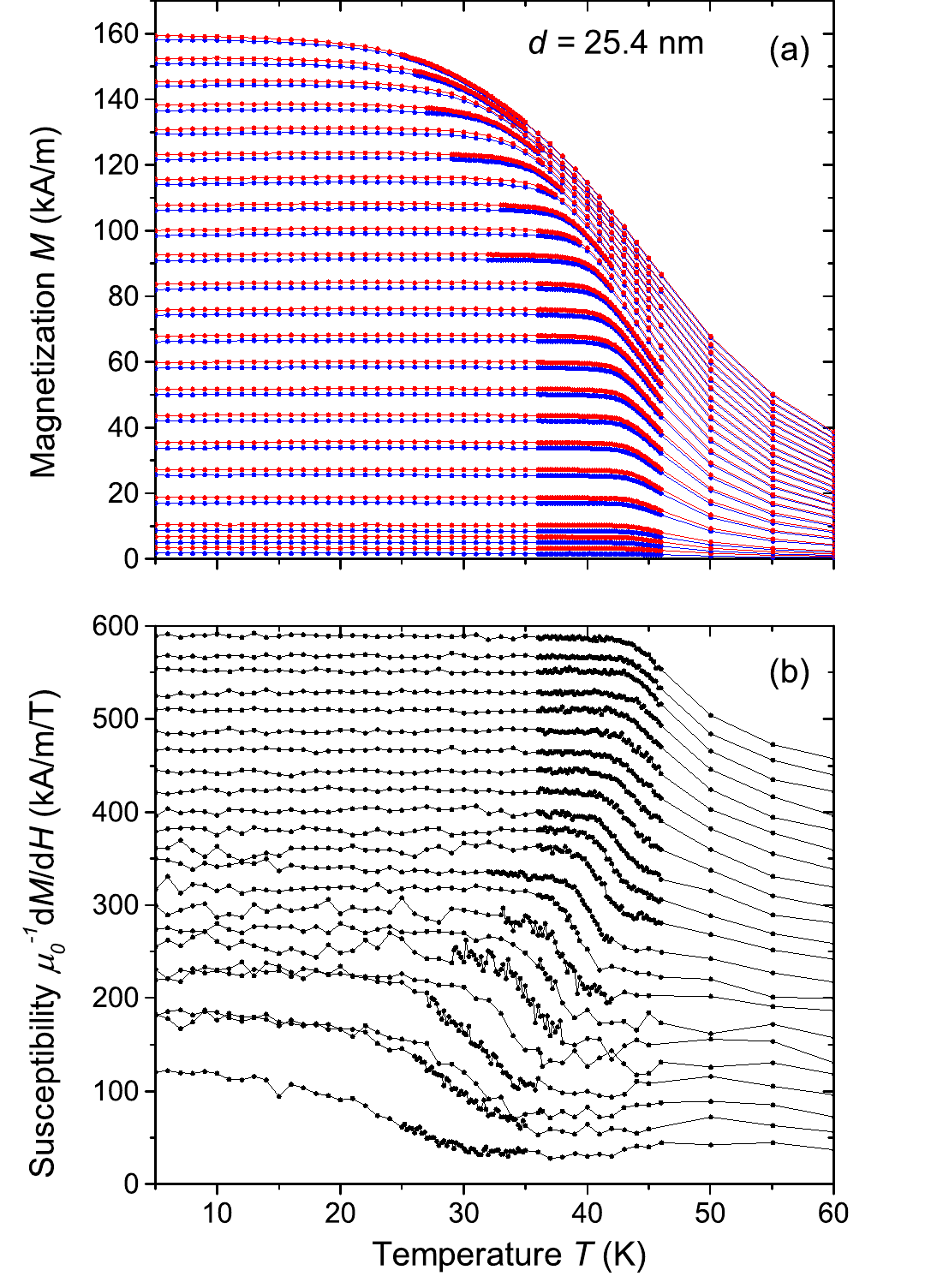}
\caption{Field cooled magnetization measurements for a $d=25.4$~nm MnSi/Si(111) film with $\mathbf{H}\| [111]$. (a) Data sets shown in blue are for field values in steps of 0.05~T from 0.05 to 0.6~T and steps of 0.1~T from 0.7~T to 1.0~T, and the data sets in red are each measured at a field 10~mT higher than the blue. (b) Field cooled static susceptibility  $dM/dH$ calculated from the pairs of red and blue $M(T)$-curves. Field values shown are in steps of 0.05~T from 0.055 to 0.605~T and steps of 0.1~T from 0.705~T to 1.005~T from top to bottom, and are separated by 20~kA/m/T for clarity.}
\label{fig:FC}
\end{figure}

From $M-H$ scans, we calculated the static susceptibility, $dM/dH$, as a function of both temperature and field in order to search for any indication of a magnetic phase transition below the saturation field $H_{C2}$. In Fig.~\ref{fig:hys}, we present the measured $M-H$ curves obtained between 5 K and 50 K, which are qualitatively similar to what is found in bulk.\cite{Bloch:1975pla} We present each of these curves on only a single branch, alternating increasing and decreasing field, as we saw no hysteresis in the full hysteresis loops taken over several temperatures between 5 K and $T_C$.

Unlike the case for bulk MnSi samples \cite{Kadowaki:1982jpsj} and MnSi epilayers with in-plane magnetic fields,\cite{Wilson:2012prb} there are no peaks in the $dM/dH$ of Fig.~\ref{fig:hys}(b) that would signal the existence of chiral modulations other than the cone phase. The only magnetic phase transition that is visible is the second order transition delineated by the inflection point in the $dM/dH$-curves at a field $H_{sat}^\perp$ that we attribute to the onset of the saturated state at $H_{C2}$.  We present the temperature dependence of $H_{C2}^\perp$ in Fig.~\ref{fig:Hc2}. For fields near 0.1~T, $dM/dH$ drops for all scans measured below $T_C$. We attribute this to a small sample misalignment in the straw used to hold samples for SQUID measurements which mixes in a small amount of the uncompensated in-plane magnetic moment into the out-of-plane $M-H$ measurements.  We confirm the absence of hysteresis in the out-of-plane $M-H$ loops by MR measurements presented in the next section.

To screen for first-order transitions that may have phase boundaries along a vertical line on a $T-H$ plot, we calculated the static susceptibility from field-cooled magnetization measurements.  For an in-plane magnetic field, such measurements produced clear peaks in $dM/dH$ at the skyrmion phase boundary of MnSi thin films,\cite{Wilson:2012prb} and measurements of bulk samples have produced peaks corresponding to the transition in and out of the $A$-phase.\cite{Kadowaki:1982jpsj, Bauer:2012prb} Samples were cooled in a fixed magnetic field from $T=100$~K to 5~K, and the magnetization was measured on warming. The curves from these measurements are shown in Fig.~\ref{fig:FC}.  In Fig.~\ref{fig:FC}(b) we constructed $dM/dH$-curves from the pairs of data sets separated by $H=10~mT$ in Fig.~\ref{fig:FC}(a). These figures show no peaks that would indicate the presence of a first-order magnetic phase transition.

\subsection{Magnetoresistance}
The existence of the $A$-phase pocket is also observable in magnetoresistance (MR) measurements, as shown by Kadowaki \emph{et al.}\cite{Kadowaki:1982jpsj} who observe hysteretic peaks in the MR near the $A$-phase boundary.  
We use MR measurements as further evidence of the absence of first-order magnetic phase transitions in MnSi/Si(111) in out-of-plane magnetic fields, and to probe the magnetic phase diagram with a higher density of field-temperature points.
While such features are present at the skyrmion-helicoid boundaries in MnSi thin films for in-plane magnetic fields, the out-of-plane MR do not show such trademarks.

\begin{figure}[h!]
\centering
\includegraphics[width=87 mm]{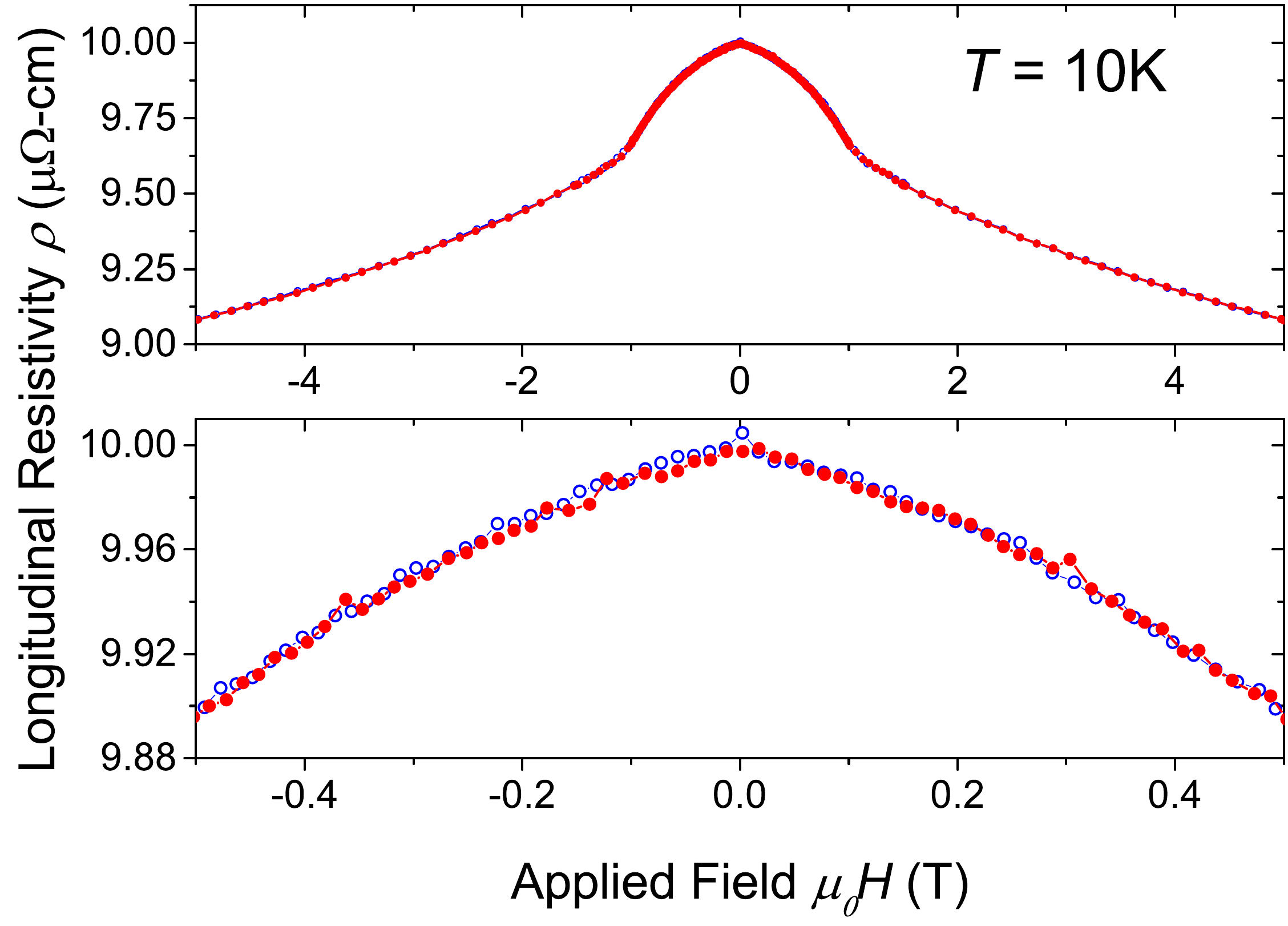}
\caption{Resistivity of a 25.4-nm thick MnSi/Si(111) film at $T=10$~K, $H\|[111]$, for increasing (red filled circles) and decreasing fields (blue open circles). No hysteresis is observed in this data or at any other temperature.
}
\label{fig:rhohys}
\end{figure}

\begin{figure}[h!]
\centering
\includegraphics[width=78 mm]{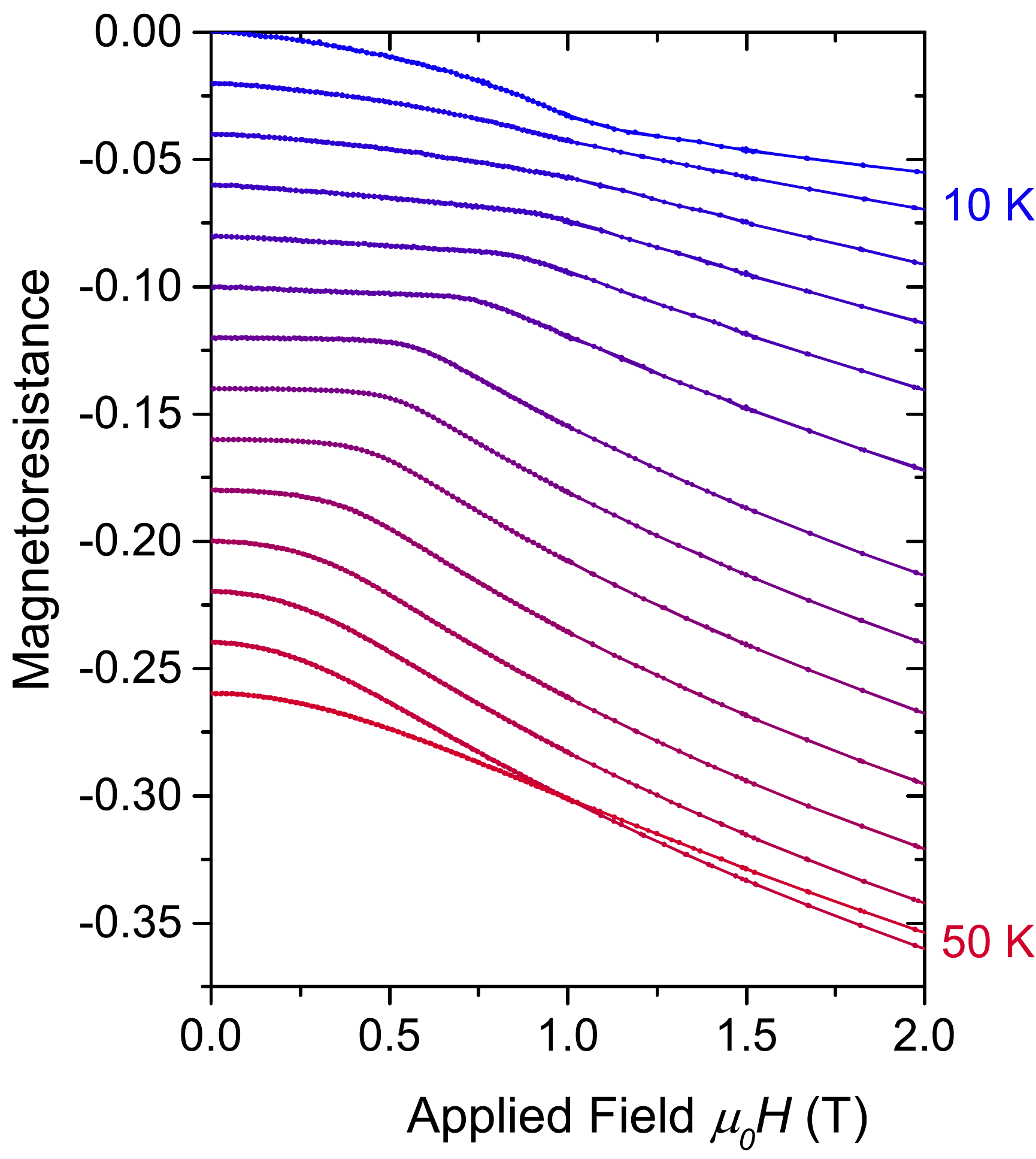}
\caption{Magnetoresistance $(\rho(H) - \rho(0))/\rho(0)$ of a 25.4-nm MnSi/Si(111) film with $\mathbf{H}\| [111]$. Temperatures shown are 10, 15, 20, 25, 30, 35, 40, 41, 42, 43, 44, 45, and 46K. Curves are offset by 0.015 for clarity.
}
\label{fig:drho}
\end{figure}

Fig. \ref{fig:rhohys} shows representative resistivity curves for both increasing and decreasing field scans measured at $T=10$~K. The sample mount used for MR measurements allowed very accurate sample alignment perpendicular to the applied magnetic field, in contrast to the SQUID measurements where the alignment is only accurate to within a few degrees. The magnetoresistance data is thus a more reliable indicator of the true out-of-plane hysteresis of the sample, and the lack of any hysteresis in this configuration is supporting evidence that the drop in $dM/dH$ observed at 0.1~T in Fig. \ref{fig:hys} is not intrinsic, but is rather due to a small sample misalignment. 
Furthermore, the lack of hysteresis or peaks in the MR is additional evidence for the absence of the $A$-phase pocket in this sample.  The temperature dependence of the MR =$(\rho(H) - \rho(0))/\rho(0)$ in Fig. \ref{fig:drho} demonstrates that it varies smoothly over all fields and temperatures and supports the conclusion that no skyrmions exist in out-of-plane magnetic fields.

\subsection{Phase diagram of MnSi/Si(111) for $\mathbf{H}\|[111]$}

\begin{figure*}
\centering
\includegraphics[width= 174 mm]{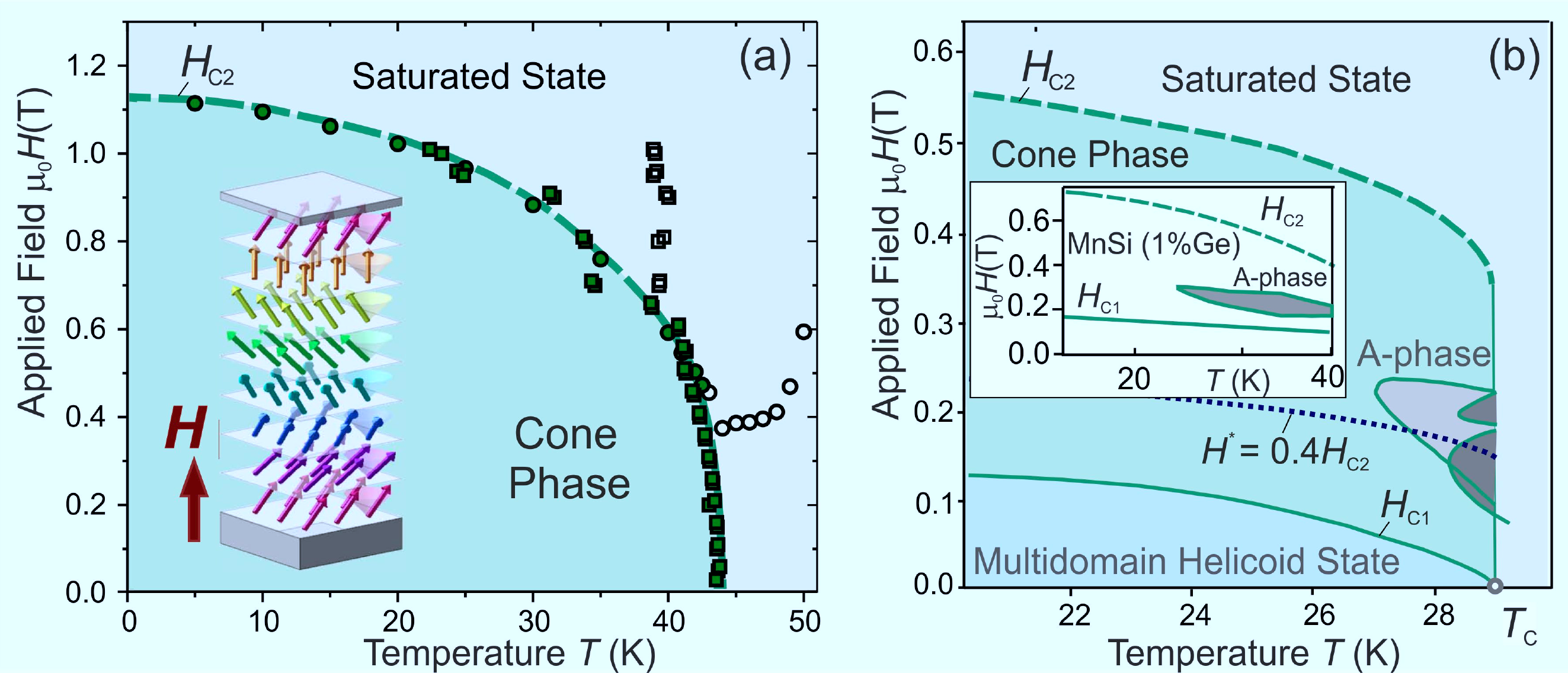}
\caption{
(a) The phase diagram of a $d =25.4$~nm MnSi/Si(111) film with $\mathbf{H}\|[111]$.
The filled circles (squares) show transition fields obtained from the minima in $d^2M/dH^2$ ($d^2M/dT^2$) calculated from the data in Fig.~\ref{fig:hys} (Fig.~\ref{fig:FC}).  Only a single phase boundary is seen, the boundary between the conical and ferromagnetic phases.  The area with multi-domain helicoid states and the $A$-pocket, characteristic of bulk MnSi and other cubic helimagnets (b), are suppressed by the strong hard-axis uniaxial anisotropy.   In addition, transition fields in between the saturated state and paramagnetic states are shown by the open circles and squares.
(b) The $T-H$ phase diagrams of bulk MnSi near $T_{C}$ \cite{Kusaka:1976ssc, Ishikawa:1977prb} is compared to a Ge-doped MnSi crystal in the inset (b).\cite{Potapova:2012prb} Along the dotted line $H^* (T)$, the difference between the energy densities of the skyrmion lattice and the cone phase is minimal (see inset in Fig. \ref{fig:phasediagram}).
} 
\label{fig:Hc2}
\end{figure*}

We summarize the experimental results of Figs. \ref{fig:hys}, \ref{fig:FC}, \ref{fig:rhohys} and \ref{fig:drho} with a  construction of the magnetic phase diagram for MnSi/Si(111) films in terms of temperature and the perpendicular applied field (Fig. \ref{fig:Hc2}(a)). 
The cone phase and the field-induced ferromagnetic state are the only thermodynamically stable states below the ordering temperature.
 The critical fields $H_{C2} (T)$ obtained from the minima in $d^2M/dH^2$  separate these two regions and are consistent with the features in the MR data in Fig.~\ref{fig:drho}. Above the Curie temperature, the minimum in $d^2M/dH^2$ persists and is shown by the open circles in Fig.~\ref{fig:Hc2}(a). In addition, there is a weak feature visible in the $dM/dH$ data of Fig.~\ref{fig:FC}(b) at higher fields.  These two features are representative of the broad cross-over region between the field induced saturated state and the paramagnetic state, as observed in bulk MnSi,\cite{Ishikawa:1977prb,Thessieu:1997jpcm, Demishev:2012prb} and FeGe.\cite{Wilhelm:2011prl} 

We obtain additional confirmation of $H_{C2} (T)$ from the inflection point in the $dM/dH$ data of Fig.~\ref{fig:FC}(b).  While the critical fields obtained from this method are consistent with the $H_{C2} (T)$ values from Fig.~\ref{fig:hys}(b), the $M(T)$ data were less noisy.  We therefore used the minima in $d^2M/dT^2$ as a measure of the temperatures of the phase transition at a given value of $H$.  For higher fields, a second minimum is present in $d^2M/dT^2$ due to the cross-over region and is shown by the open squares in Fig.~\ref{fig:Hc2}(a).

The magnetic phase diagram for MnSi epilayers (Fig. \ref{fig:Hc2} (a)) differs from the corresponding phase diagram for bulk cubic helimagnets (Fig. \ref{fig:Hc2}(b)). The region with multi-domain helicoid states bounded by the critical line $H_{C1} (T)$ and a tiny closed area near the Curie temperature, namely the $A$-phase pocket that exists bulk MnSi,\cite{Kusaka:1976ssc, Ishikawa:1977prb} both disappear from the phase diagram of MnSi/Si(111) epilayers. 
In Section \ref{sec:dis}, we discuss the physical mechanisms underlying the formation of these areas in bulk cubic helimagnets and their modification in confined samples.

\section{Discussion}
\label{sec:dis}

\subsection{Absence of $\mathbf{H_{C1}}$ in MnSi/Si(111)}
Multi-domain helicoid states arise as a result of the degenerate ground state in bulk cubic helimagnets.   At zero field, the helices propagate along the directions imposed by cubic anisotropy, which are the $\langle111\rangle$ directions in the case of MnSi. 
The applied magnetic field lifts the degeneracy of these propagation directions, selects the one along the direction of the applied field, and transforms the helix into the cone phase.  In bulk cubic helimagnets a magnetic-field induced reorientation of the helices develops a complex process including a displacement of the domain boundaries and a rotation of the propagation directions within the domains.\cite{Plumer:1981jpc, Grigoriev:2006prb1} These processes are similar to magnetic-field induced transformations of multi-domain states observed in many classes of magnetically ordered materials and are decribed by common micromagnetic equations.\cite{Hubert:1998, Baryakhtar:1988spu}  In cubic helimagnets, the reorientation of the helicoids ends at the critical field $H_{C1} (T)$  with the formation of a single-domain cone phase. 

In epitaxial MnSi nano-layers, a strong hard-axis anisotropy favors helices with the propagation direction perpendicular to the film surfaces and suppresses helices with other propagation directions.
As a result, the propagation direction is homogeneous in the ground state of such films. 
Due to the existence of inversion domains in the crystal structure,\cite{Karhu:2010prb} there are variations in the magnetic chirality on the length scale of the order of 1~$\mu$m.\cite{Karhu:2011prb}  The magnetic frustration between these regions creates magnetic domains that display a glassy-magnetic behavior for fields applied in-plane.\cite{Karhu:2010prb}   Nevertheless, the uniformity of the propagation direction explains the absence of the multi-domain helicoids in out-of-plane magnetic fields  and the lack of a critical line $H_{C1} (T)$ in the magnetic phase diagram of MnSi films. 
Results that are consistent with these facts have been reported for 9 nm and 19 nm thick epitaxial MnSi films by others.~\cite{Menzel:2013jkps}

\subsection{ Nature of skyrmions in MnSi nano-layers}
\label{sec:suppr}

The strong uniaxial anisotropy that arises in epitaxial films of cubic helimagnets drastically changes the energy balance between the various modulated states, as shown in Fig. \ref{fig:phasediagram}. 
The strong $K>0$ anisotropy observed in all FeGe/Si(111) films (inset of Fig.~\ref{fig:phasediagram})~\cite{Huang:2012prl} lies within the $[K_B, K_C]$ interval of the calculated $K-H$ phase diagram in Fig.~\ref{fig:phasediagram}.  Huang \emph{et al.}\cite{Huang:2012prl} report the existence of skyrmion lattices in a range of magnetic fields that are in agreement with the calculated critical fields $H_1$ and $H_s$ for the skyrmions lattice in Fig.~\ref{fig:phasediagram}.  The calculations show that in easy-axis FeGe films, the uniaxial anisotropy effectively suppresses the cone phase and stabilizes the skyrmion lattice in a broad range of the applied fields and temperatures. 
In contrast, MnSi/Si(111) epilayers exhibit a strong hard-axis uniaxial anisotropy. Figure~\ref{fig:phasediagram} shows the range of anisotropies spanned by the MnSi films in Ref.~\onlinecite{Karhu:2012prb}. In such nano-layers, elliptically distorted skyrmions have been found to exist in a broad range of in-plane magnetic fields.\cite{Wilson:2012prb} For a perpendicular magnetic field on the other hand, the hard-axis uniaxial anisotropy ($K<0$) in MnSi/Si(111) shifts the energy balance in favor of the cone phase (lower panel of Fig. \ref{fig:solutions2}).  As a result, the $K<0$ entirely suppresses the formation of a helicoid and a skyrmion lattice with in-plane propagation directions (Fig. \ref{fig:phases} (a) and (c)).   
These theoretical results, supported by the experimental results in this paper and by numerous others,\cite{Karhu:2010prb,Karhu:2011prb,Karhu:2012prb, Wilson:2013prb, Menzel:2013jkps} exclude the existence of in-plane helicoids and (111)-oriented skyrmions (Fig. \ref{fig:phases}) in hard-axis MnSi/Si(111) epilayers and establish that a helix with a propagation direction along (111) is the only magnetic ground state.

These findings, however, have been recently disputed in Ref.~\onlinecite{Li:2013prl}. Based on Lorenz microscopy measurements of a 10-nm thick MnSi/Si(111) epilayer, the authors of Ref.~\onlinecite{Li:2013prl} claim that in-plane helicoids and skyrmions lattices exist in a broad range of out-of-plane magnetic fields in contrast to the theoretical results summarized in Fig.~\ref{fig:phasediagram}, which show that these states would only be present for $K>0$.  Loudon recently demonstrated that the contrast in the Lorentz images published by Li \emph{et al.} are due to structural artifacts and are not of magnetic origin: the same features observed in Ref.~\onlinecite{Li:2013prl} are also observed at room temperature, far above $T_C$.\cite{Monchesky:2013prl}  It is important to point out that the interpretation put forward in Ref.~\onlinecite{Li:2013prl} is not only at odds with theoretical calculations, but it also contradicts the experimentally established facts.  The striped pattern in Fig.~1(a) in Ref.~\onlinecite{Li:2013prl} is incorrectly interpreted as an in-plane helicoidal ground state.  In-plane helicoids would have zero remanent magnetization, contrary to what is reported by others.\cite{Magnano:2010apl, Karhu:2010prb} MnSi/Si(111) films clearly show oscillations in the remanent magnetization with thickness with a wavelength given by the pitch of the helix that rules out the existence of in-plane helicoids.\cite{Karhu:2011prb}  Furthermore, PNR conclusively shows that the propagation vector of the helix points out-of-plane.\cite{Karhu:2011prb, Karhu:2012prb, Wilson:2013prb}  The striped pattern is, however, explained by moire fringes and is perfect agreement with the strain reported in Ref.~\onlinecite{Karhu:2012prb}.

\subsection{Why the $A$-phase exists}
\label{sec:aphase}

The suppression of an $A$-phase pocket in MnSi/Si(111) near the ordering temperature in Fig.~\ref{fig:Hc2} due to $K<0$ provides further evidence for the delicate energy balance that exists in the $A$-phase in bulk MnSi crystals.  Analysis of the magnetic-field-driven evolution of the skyrmion lattice period and the energy $\Delta E_{h} (H/H_D)$ (Fig.~\ref{fig:solutions2}) allows one to understand the physical mechanism that leads to the formation of the $A$-phase.
For $K = 0$ (bulk helimagnets), the cone phase is the global minimum of the functional $w_0$ over the whole magnetic field range where modulated states exist ( $0 < H < H_D$), as indicated by the fact that both $\Delta E_{h}$ and $\Delta E_{s}$ are always positive.  
However, two-dimensional chiral modulations provide a larger reduction of the DM interaction energy in skyrmion lattices compared to one-dimensional helical modulations.   Calculations within the model of Eq.~(\ref{density}) show that this reduction increases with increasing field up to a field $H^*$ as the equilibrium sizes of the skyrmion cell decreases.\cite{Bogdanov:1994jmmm}  At $H^*$, the skyrmion lattice reaches its highest density and lowest energy difference $\Delta E_{s} (H^*) \equiv \mathrm{min} [\Delta E_{s}] $, as seen by a comparison between the $K=0$ line in the lower panel of Fig.~\ref{fig:solutions2} to the point (a) in the upper panel.  At higher fields $ H^* < H < H_s$ the skyrmion lattice gradually expands into a set of isolated skyrmions at a critical field $H_s = 0.813 H_D$.\cite{Bogdanov:1994jmmm} 

Nevertheless, skyrmion lattices are only metastable for $K=0$ 
and additional interactions are required to stabilize them.  
The size of the anisotropy given by Eq.~\ref{cubic}
gradually decreases as  $\mathrm{min} [\Delta E_{s} (T)] $ decreases with increasing $T$ and becomes zero at $T_C$.
This means that even small perturbations can suppress the cone phase and lead to the formation of a skyrmion lattice in a pocket about the $H^*(T)$ line
(see Fig. 7 in Ref.~\onlinecite{Wilhelm:2012jpcm}). 
The small size of the energy imbalance and low potential barriers that characterize this region make the $A$-phase pocket extremely sensitive to small interactions, such as the softening of the magnetization modulus, dipolar interactions, fluctuations, and anisotropy.   In particular, calculations show that an exchange anisotropy as small as $ B = 0.1 K_0$ (Eq.~(\ref{cubic})) is sufficient to create a thermodynamically stable skyrmion lattice in a certain field range near $H^*$.\cite{Butenko:2010prb}  The importance of this anisotropy is evidenced by a number of experimental results, including the variation in the size of the $A$-phase pocket with the orientation of the magnetic field,\cite{Lamago:2006pb} and the increase in the size of the $A$-phase region in MnSi by doping with a larger spin-orbit interaction element.\cite{Potapova:2012prb} 

The behavior in bulk crystals contrasts the behavior in epilayers discussed in Section~\ref{sec:suppr}, where a uniaxial anisotropy dominates the small interactions discussed above, and either suppresses the $A$-phase entirely, or results in the stabilization of a skyrmion lattice over large regions of the phase diagram.  To explore the evolution of the skyrmion phase between the behaviors observed in bulk and those in films, it would be interesting to investigate the influence of a uniaxial pressure on bulk cubic helimagnets.

The sensitivity of chiral modulations in the $A$-phase to weak interactions leads to several complex magnetic states as observed in many cubic chiral helimagnets.\cite{Kadowaki:1982jpsj, Wilhelm:2011prl, Seki:2012prb1, Moskvin:2013prl}   This complex behavior is demonstrated theoretically when a soft magnetization modulus is included in the calculation.\cite{Wilhelm:2011prl}  However, the exact structure of the complex magnetic textures reported in Refs.~\onlinecite{Lamago:2006pb, Grigoriev:2006prb2, Stishov:2008jetp,  Muhlbauer:2009sci, Pappas:2009prl, Lee:2009prl, Wilhelm:2011prl, Onose:2012prl, Wilhelm:2012jpcm, Bauer:2012prb, Potapova:2012prb, Moskvin:2013prl} and the particular physical mechanism underlying the stabilization of these states are still unresolved and remains the subject of controversy between different research groups.\cite{Rossler:2006nat, Binz:2006prl, Grigoriev:2006prb2, Grigoriev:2010prb, Ho:2010prb, Wilhelm:2011prl, Moskvin:2013prl} 
(For details see  review papers \cite{Rossler:2011jpcs,Wilhelm:2012jpcm}).
Our results on the precursor states evolution in confined cubic helimagnets (and, particularly, the conclusion about the suppression of the A-phase pocket in easy-plane  MnSi/Si (111) epilayers) are based on the analysis of the general energy
balance between the competing cone phase and the multidimensional magnetic modulations in the A-phase, but do not depend on the specific details of the textures in this problematic region.

\begin{figure}[!]
\includegraphics[width = 87 mm]{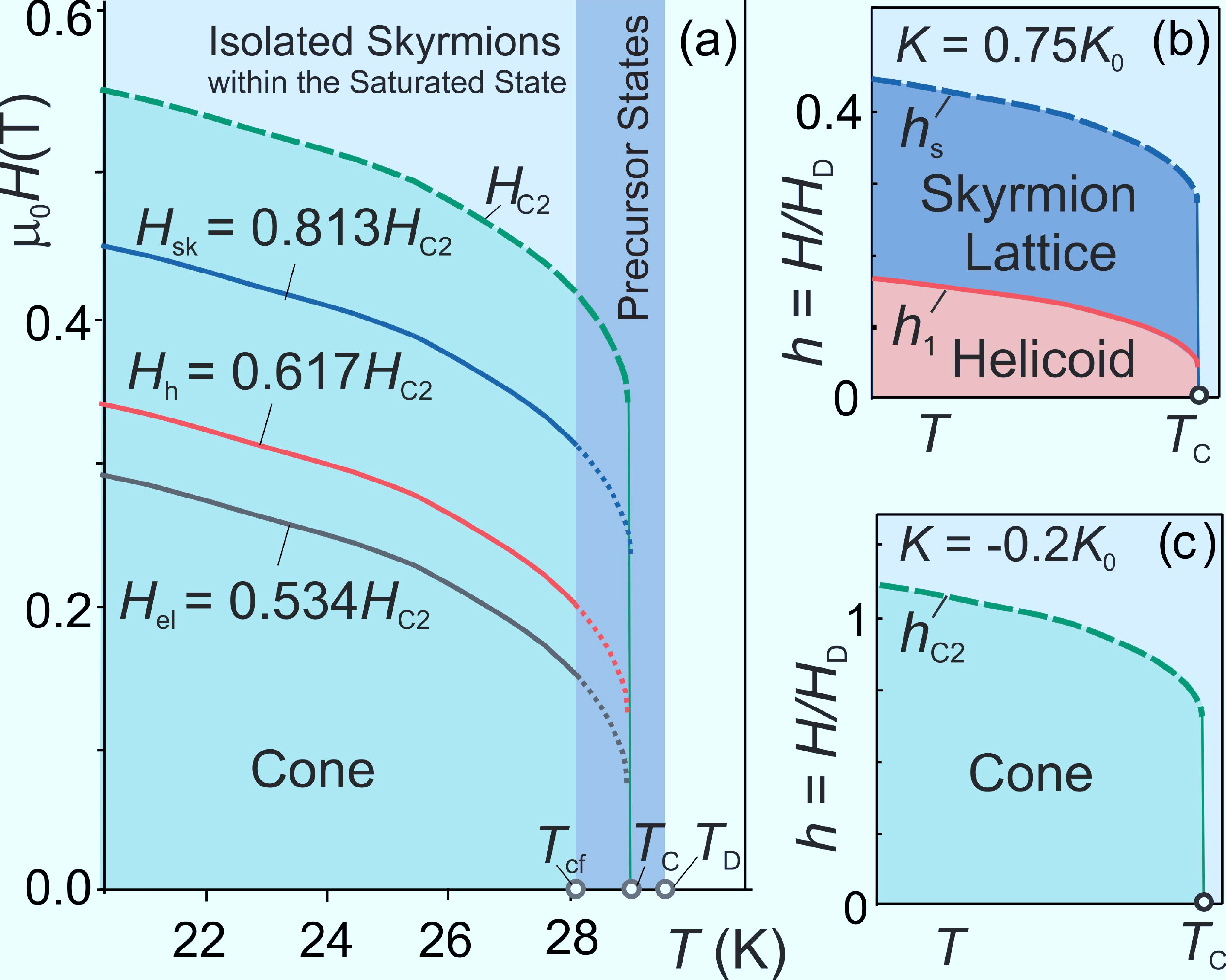}
\caption{
Calculated $T-H$ phase diagram for a bulk MnSi  based on the solutions for $K = 0$ in Fig. \ref{fig:phasediagram}. Thin lines bound  the existing areas for the helicoids ($H_h (T) $), the skyrmion lattice ($H_{sk} (T) $), isolated skyrmions ($H_{el} (T) $); $H_{1}$ is the line of the phase equilibrium between the metastable helicoid and skyrmion lattice. Two characteristic temperatures, \textit{confinement} temperature $T_{cf}$ and \textit{nucleation} temperature $T_D$ bound the precursor region ($ T_{cf} < T < T_D$)  \cite{Wilhelm:2012jpcm}.Typical $T-H$ phase diagrams for cubic helimagnets with easy-axis   (b)  and easy-plane (c) uniaxial distortions.
 }
\label{fig:THtypes}
\end{figure}

\subsection{$\mathbf{T-H}$ phase diagrams of bulk crystals revisited}

\label{sec:revisit}
The  $T-H$ phase diagram presented in Fig. \ref{fig:Hc2}(b) was constructed from the first papers dedicated to magnetic properties of MnSi \cite{Kusaka:1976ssc, Ishikawa:1977prb} and other cubic helimagnets.\cite{Ishimoto:1995pb} This diagram was later explained through several theoretical and experimental efforts (see, e.g., Refs.~\onlinecite{Plumer:1981jpc,Lebech:1989jpcm}). 
However, recent progress in our understanding of the magnetization 
processes now enables us to update the `canonical' magnetic phase diagram of cubic helimagnets to include regions of metastability and the precursor region, both of which are necessary to understand the collection of measurements of these materials.
With this purpose, we address here a problem of metastable states and discuss a `hierarchy' of magnetic states arising in cubic helimagnets.
Nearly all representations of the phase diagram consider only the equilibrium phases and ignore the metastable states.   In Section~\ref{sec:theory}, however, we present regions of stability and metastability. Metastable states are important in magnetization processes in general, as described in Ref.~\onlinecite{Hubert:1998}.  This is true for the first-order transitions between states with different topology.  Metastable states are seen in the regions of mixed phase in (Fe,Co)Si nano-layers,\cite{Yu:2010nat} and MnSi thin films.\cite{Wilson:2012prb}  More recently, field cooling experiments in bulk Fe$_{0.5}$Co$_{0.5}$Si managed to form a metastable skyrmion lattice.\cite{Milde:2013sci} The isolated skyrmions reported in Refs.~\onlinecite{Yu:2010nat, Romming:2013sci} are another example of metastability.    To facilitate a comparison between theory and experiments, we produce a  theoretical $T-H$ phase diagram in Fig.~\ref{fig:THtypes}(a) that includes regions of metastability based on the results of Section~\ref{sec:theory}. 

In a broad temperature range, the magnetization modulus is practically uniform over the material but has a temperature dependence  $\mathbf{M} (T) = \mathbf{M} (0) \sigma (T)$.
The temperature dependence in Fig.~\ref{fig:THtypes} is obtained from solutions to Eq.~(\ref{density}) by using the reduced magnetization $\sigma(T)$ and $H_{C2}(T)$ for bulk MnSi. In Fig.~\ref{fig:THtypes}(a) we use the calculated critical fields of the modulated states for $K = 0$ (Fig.~\ref{fig:phasediagram}) to obtain the theoretical equilibrium phase boundaries for the cone phase ($H < H_{C2}(0) \sigma (T))$, the metastable helicoid ($H < H_{h} \sigma (T))$, skyrmion lattice ($H < H_{s} \sigma (T))$, and for isolated skyrmions ($ H > H_{el}\sigma (T)$).  This provides a good description of the MnSi phase diagram over most of the phase diagram.

The updated $T-H$ phase diagram for MnSi in Fig.~\ref{fig:THtypes} provides a framework in which to understand the field cooling experiments in bulk Fe$_{0.5}$Co$_{0.5}$Si.\cite{Milde:2013sci}  By cooling through the precursor region at an appropriate field below $ H_{sk} = 0.813 H_{C2}$,  stable skyrmion lattices are nucleated in this region characterized by low energy barriers.  As the temperature drops below the precursor region, the skyrmion lattices become metastable.  However, the barrier heights increase with decreasing temperature and provide the robustness of these metastable states.  When the field is then reduced at fixed $T$, $H_{el}$ is eventually reached where the skyrmions strip-out into helicoids, as observed in Ref.~\onlinecite{Milde:2013sci}.

The temperature dependence for $K \ne 0$ is shown in Fig.~\ref{fig:THtypes}(b) and (c) by again using the results of Section~\ref{sec:theory} with $\sigma(T)$ for bulk MnSi.  These figures facilitate a comparison with the thin film experiments.  Figure~\ref{fig:THtypes}(b) captures well the qualitative behavior of MnSi in Fig.~\ref{fig:Hc2}, while Fig.~\ref{fig:THtypes}(c) is able to reproduce the stable skrymion and helicoid regions of FeGe/Si(111) in Ref.~\onlinecite{Huang:2012prl}.
Spatial modulations of the magnetization modulus, while negligible in a broad temperature range, become a sizeable effect in the vicinity of the ordering temperature.\cite{Leonov:2010arXiv, Wilhelm:2012jpcm}
In this precursor region, the magnetic textures also display spatial variations of the magnetization modulus, which strongly modifies their properties compared  to \textit{regular} modulations that arise at lower temperatures.
A theoretical treatment within the Dzyaloshinskii-Bak-Jensen model that accounts for this additional degree of freedom reveals two characteristic temperatures that define the precursor region, namely the \textit{confinement} temperature, $T_{cf} = T_0 -3D^2/(4 JA)$ and the \textit{nucleation} temperature, $T_D = T_0 +D^2/(2JA)$ shown in Fig.~\ref{fig:THtypes}(a).\cite{Rossler:2006nat, Wilhelm:2011prl}  These separate the precursor region from the paramagnetic phase on one side from the region with regular chiral modulations on the other.
 The peculiarities of the $T-H$ phase diagram in this region are discussed in Ref.~\onlinecite{Wilhelm:2012jpcm}.

Finally, we note that basic chiral modulations shown in Fig. \ref{fig:phases} arise in cubic helimagnets as a result of competition between the main magnetic interactions and are described by regular solutions of the Dzyaloshinskii-Bak-Jensen model. 
These should not be confused with ``weak'' magnetic states in the $A$-pocket where a clear hierarchy of interactions disappears and the energy barriers that protect the states are small.

\subsection{Surface effects in chiral ferromagnets} 

Recent experimental and theoretical findings demonstrate that surface effects may stabilize specific chiral modulations in confined cubic helimagnets as skyrmions modulated along three spatial directions,\cite{ Rybakov:2013prb} or twisted states at high in-plane fields.\cite{Wilson:2013prb}
Theoretical analysis shows that the DM interactions near the surfaces of cubic helimagnets induce specific chiral modulations with the propagation direction perpendicular to the sample surfaces  
(\textit{chiral twists}) \cite{Wilson:2013prb, Rybakov:2013prb}.
In chiral helimagnets films where $d \leq L_D$, such surface twists become a sizeable effect and strongly modify the skyrmion energetics and provide a thermodynamic stability to the skyrmion lattice in a broad range of applied magnetic fields.\cite{Rybakov:2013prb} These results elucidate recent observations of skyrmion lattices in free-standing cubic helimagnets nano-layers (see e.g Ref.~\onlinecite{Yu:2011nm}), whereas skyrmions are suppressed by one-dimensional (conical) modulations in bulk crystals of the same material.
It was also established by numerical calculations that similar surface modulation instabilities strongly influence the structure of isolated skyrmions in magnetic nanodots. \cite{Rohart:2013prb, Sampaio:2013nn}
 Furthermore, we note that in MnSi/Si (111) epilayers, surface effects compete with the in-plane uniaxial anisotropy.\cite{Rybakov:unpub} However, for the $d = 25.4~\textrm{nm} = 1.83 L_D$ film investigated in this paper, a strong in-plane anisotropy sufficiently weakens the finite size effects in a broad range of magnetic fields and temperatures.

\section{Materials and perspectives}
\label{sec:mat}

In this section we briefly overview the existing groups of confined noncentrosymmetric magnets  and discuss how the induced and intrisic uniaxial anisotropy influences chiral modulations in these compounds. 
\\

\subsection{Free-standing films of cubic helimagnets.} 
The first images of chiral skyrmions have been observed in
20-nm thick mechanically thinned B20 Fe$_{0.5}$Co$_{0.5}$Si samples. \cite{Yu:2010nat}
Subsequently, the formation and evolution of skyrmions and helicoids were investigated in similar \textit{free-standing} layers of other B20 compound (see e.g. Refs.~\onlinecite{Yu:2011nm,Tonomura:2012nl,Yu:2013nl}). Unfortunately no $M(H)$ measurements have been carried out in these films and values of the induced uniaxial anisotropy are unknown. However, the observed magnetization processes in these compounds demonstrate features characteristic of easy-axis type of anisotropy.

The different B20 material systems display a range of behaviours in external magnetic fields that are explained with the aid of the phase diagram in Fig.~\ref{fig:phasediagram} by differences in the size of $K$.  The magnetic-field-induced evolution of magnetic states observed in (Fe,Co)Si free-standing films transforms from the \textit{helicoid $\Rightarrow$ the skyrmion lattice $\Rightarrow$  the saturated state with isolated skyrmions} (this corresponds to interval (5), $  K_{B} < K < K_{C} = 1.90$ in the Appendix).
In FeGe nano-layers \cite{Yu:2011nm} by contrast, the observed sequence of magnetic configurations follows from \textit{the helicoid $\Rightarrow$ the skyrmion lattice $\Rightarrow$ the cone phase $\Rightarrow$  the saturated state} (which is characteristic for interval (4), $  K_{A} < K < K_{B} = 0.363$ in the Appendix). 

\subsection{ Chiral helimagnetic epilayers} 

The fabrication of MnSi nano-layers on Si(111) \cite{Evans:1996prb, Shivaprasad:1997ss, Schwinge:2005jap,Higashi:2009prb,Magnano:2010apl} opened the possibility of exploring the magnetic properties of chiral thin films.\cite{Karhu:2010prb} This work has introduced a new class of nanomagnetic systems, \textit{epitaxial chiral helimagnet thin films} which are more amenable than mechanically thinned layers to the investigation of skyrmion states and other nontrivial chiral modulations with multiple techniques.\cite{Karhu:2011prb,Karhu:2012prb,Wilson:2012prb,Li:2013prl, Menzel:2013jkps}.  Investigations in other B20 epilayers, FeGe/Si,\cite{Huang:2012prl} (Fe,Co)Si,\cite{Porter:2012prb} and MnGe \cite{Engelke:2013jpcm} present a wide range of material parameters to explore.
So far, detailed measurements of the induced uniaxial anisotropy have been carried out in epitaxial MnSi 
\cite{Karhu:2010prb,Karhu:2011prb,Karhu:2012prb} and FeGe \cite{Huang:2012prl}, which span complementary ranges of $K$ (Fig. \ref{fig:phasediagram}). 
The first measurements of the magnetic anisotropy in (Fe,Co)Si/Si(111) appeared following the preparation of this manuscript.\cite{Porter:2013arxiv}  Like the case of MnSi/Si(111), these films have an out-of-plane hard-axis.  PNR measurements shows that the helical ground state of the films also have a propagation vector pointing out-of-plane.  However, the presence of hysteresis in both the in-plane and out-of-plane $M(H)$ curves indicates that the magnetic behaviour differs from that of MnSi/Si(111).  

\subsection{ Fe and FePd nano-layers.} 
 The induced interfacial DM interactions in ultra-thin layers of common magnetic metals are capable of stabilizing skyrmion lattices,\cite{Heinze:2011np} as well as isolated skyrmions in large  out-of-plane magnetic fields.\cite{Romming:2013sci} In the $K-H$ phase diagram of Fig.~\ref{fig:phasediagram}, such isolated skyrmions exist as metasable objects within the saturated phase. 
Further measurements to determine the parameters of skyrmions and to map out the phase diagram for this system will provide important comparisons with the theoretical predictions and observations in nanolayers of cubic helimagnets. 

\subsection{ Relations to bulk uniaxial helimagnets.}

In uniaxial noncentrosymmetric ferromagnets, an intrinsic uniaxial magnetic anisotropy stabilizes similar chiral modulations as those found in cubic helimagnets with an induced uniaxial anisotropy \cite{Bogdanov:1994jmmm}. For example, the chiral magnet Cr$_{1/3}$NbS$_2$ (space group P6$_3$22) develops long-range helimagnetic order below $T_C$ with a period $L_D $ = 48.0 nm.
\cite{Miyadai:1983jpsj,Togawa:2012prl, Ghimire:2012prb}
The propagation direction of the helix along the hexagonal axis indicates a hard-axis type of uniaxial anisotropy in this helimagnet. This implies that Cr$_{1/3}$NbS$_2$ should exhibit magnetic properties similar to those observed in easy-plane epitaxial films MnSi/Si(111) in Refs.~\onlinecite{Karhu:2012prb, Wilson:2012prb} and in the present paper.
Contrary to high symmetry cubic helimagnets where the DM interactions provide three equivalent propagation directions ($w_D = D \mathbf{M} \cdot \mathrm{rot} \mathbf{M}$ in Eq.~(\ref{density})), uniaxial noncentrosymmetric magnets have a DM energy that is more complex and may include several material parameters. \cite{Bogdanov:1989jetp}
Particularly, for  Cr$_{1/3}$NbS$_2$ the DM energy contribution can be written as
\begin{eqnarray}
 w_D=-D\,\mathbf{M}\cdot \mathrm{rot}\mathbf{M}
-D_1 \left(M_x \frac{\partial M_y}{\partial z}-
M_y \frac{\partial M_x}{\partial z}\right).
\label{Lifshitzd}
\end{eqnarray}
The last term in Eq.~(\ref{Lifshitzd}) imposes a difference between in-plane modulations and those along the hexagonal axis ($z$). The $K-H$ phase diagram for Cr$_{1/3}$NbS$_2$ depends on the additional material parameter $D_1/D$, and can be obtained from Fig.~\ref{fig:phasediagram} by extending ($D_1 > 0$) or shrinking ($D_1 < 0$) the cone phase region.
Peculiarities of the magnetic properties observed for in-plane fields imply the existence of skyrmionic states in this helimagnet.\cite{ Ghimire:2013prb} These findings correlate with theoretical predictions and experimental observations of elliptically distorted in-plane skyrmions in easy-plane epitaxial MnSi/Si(111) films.\cite{Wilson:2012prb}
	
In tetragonal magnets Cr$_{11}$Ge$_{19}$,\cite{Ghimire:2012prb} and Mn$_2$RhSn,\cite{Alijani:2013jap} which belong to the $\bar{4}2m$ ($D_{2d}$) chiral point group, the DM interactions only stabilize modulations propagating in the plane perpendicular to the tetragonal axis. Chiral ferromagnets of this class are suitable objects for investigations of skyrmions and helicoid states.

\section{Conclusions}

We present experimental investigations of the magnetic states in epitaxial MnSi/Si films in perpendicular magnetic fields and theoretical analysis of chiral modulations under the influence of an induced uniaxial anisotropy.
The $K-H$ phase diagram of the solution in cubic helimagnets with induced uniaxial anisotropy (Fig. \ref{fig:phasediagram})  provides an effective tool to calculate the magnetization curves and the magnetic phase diagrams in bulk and confined helimagnets (Fig. \ref{fig:THtypes}).

Our findings show that a subtle balance between the cone and the skyrmion lattice energies  (Fig. \ref{fig:solutions2}) is violated near the ordering temperature and results in the formation of a small closed area where the skyrmion lattice becomes thermodynamically stable.
We argue that the `canonical' $T-H$ phase diagram of a cubic helimagnet (Fig. \ref{fig:Hc2}(b)) includes (i) stable regions consisting of the cone phase and the saturated state that result from the strongest interactions, and (ii) regions with multi-domain helicoids and the complex modulations in the $A$-phase that are induced by much weaker forces.  The area of the phase diagram occupied by these ÒweakÓ states  can be easily modified by external and internal distortions and even be totally suppressed, as observed in MnSi/Si(111) epilayers (Fig. \ref{fig:Hc2}(a)).
We construct the updated $T-H$ phase diagram that includes both stable and metastable solutions derived within the basic model of Eq.~(\ref{density}). We show that a strong induced uniaxial anisotropy in hard-axis MnSi/Si epilayers completely suppresses the $A$-phase area and argue that uniaxial pressure applied to a bulk cubic helimagnets would provide an effective method to investigate this phenomenon.  Together with earlier findings,\cite{Bogdanov:1994jmmm, Butenko:2010prb, Karhu:2012prb, Huang:2012prl} our results create a consistent picture of uniaxial anisotropy effects arising in confined cubic helimagnets and uniaxial bulk ferromagnets.

More detailed experimental investigations are required in both induced and intrinsic chiral magnets to provide a comprehensive picture of how anisotropy affects the magnetic properties of skyrmions and other chiral modulations and to observe the complete range of behavior predicted by the phase diagram in Fig.~\ref{fig:phasediagram}. From the theoretical side, finite-size effects indicated in Ref.~\onlinecite{Rybakov:2013prb, Wilson:2013prb} should be thoroughly investigated to complete the theoretical description of confined chiral modulations within basic model of Eq.~(\ref{density}).

\acknowledgements{
We would like to thank Filipp Rybakov, James Loudon and Mike Robertson for helpful discussions. ABB acknowledges financial support from the Deutsche Forschungsgemeinschaft (SFB 668) and the Hamburg Cluster of Excellence NANOSPINTRONICS. TLM and MNW acknowledge support from NSERC, and the support of the Canada Foundation for Innovation, the Atlantic Innovation Fund, and other partners which fund the Facilities for Materials Characterization, managed by the Institute for Research In Materials.  We would also like to thank Eric Karhu, Mike Johnson and Simon Meynell for technical assistance.}

\appendix*
\section{K-H phase diagram details}
\begin{figure}[!ht]
\includegraphics[width=87mm]{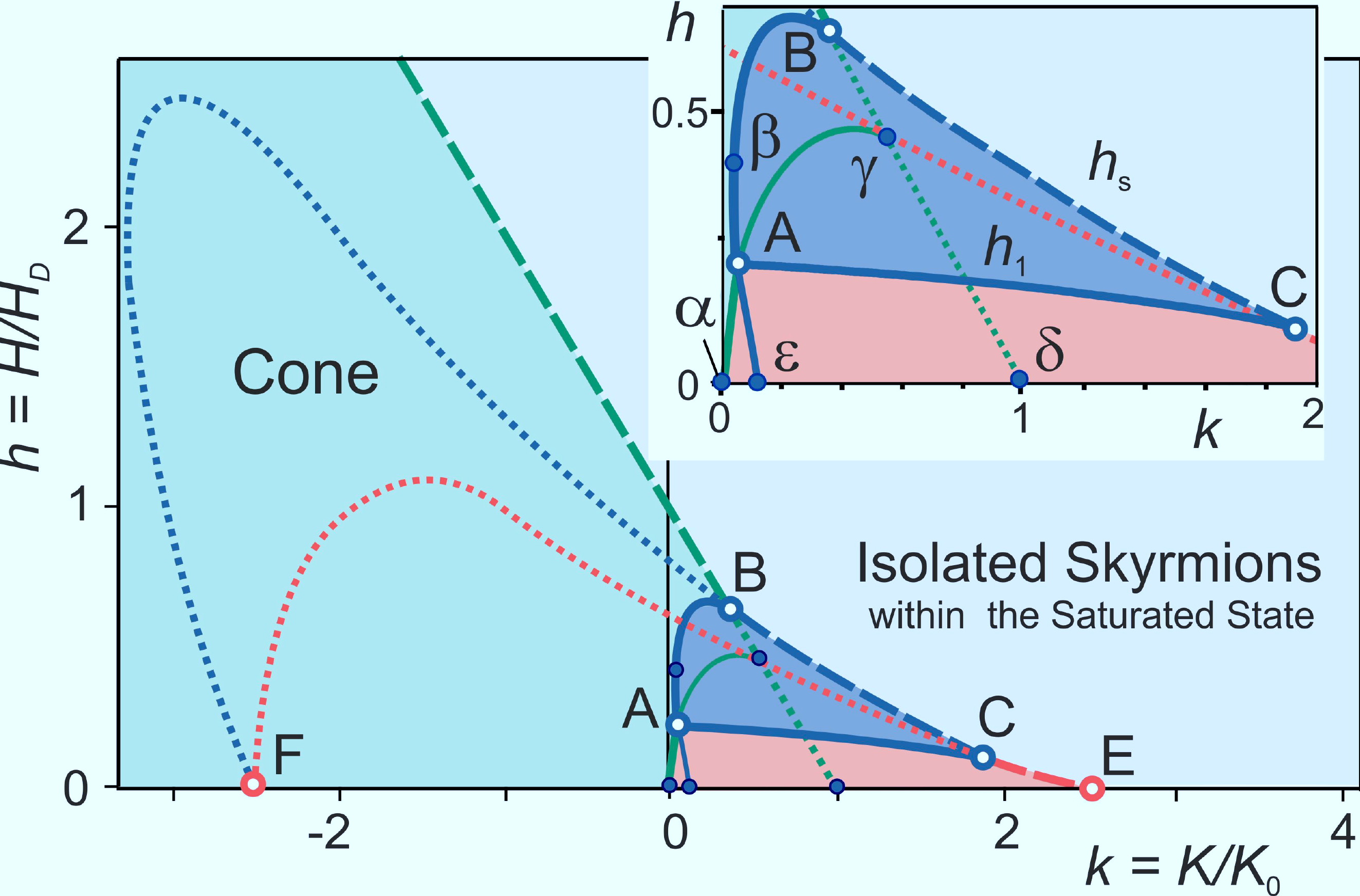}
\caption{$K-H$ phase diagram in a broad range of  induced anisotropy
$K$ and perpendicular field $H$ includes the  complete existence areas of
the (meta)stable modulated states. Inset shows the detailed phase diagram within the stability area of the skyrmion lattice.
}
\label{fig:app1}
\end{figure}

We present in Fig.~\ref{fig:app1} the $K-H$ phase diagram over a larger range of $K$ than presented in Fig.~\ref{fig:phasediagram} and collect the coordinates of the critical and characteristic points in Table 1.
 Depending on values of $K$, the $H-K$ phase diagram indicates seven different types of the magnetization curves and $T-H$ phase diagrams:

\begin{enumerate}
\item $K \leq 0$. In this region the cone phase corresponds to the global minimum of functional $w_0$ in Eq.~(\ref{density}) in the whole region where the modulated states exist. This describes magnetization processes in bulk ($K = 0$) and confined cubic helimagnets with hard-axis uniaxial anisotropy  ($K < 0$), e.g. epitaxial MnSi/Si(111) layers (Figs.~\ref{fig:hys}, \ref{fig:FC} and \ref{fig:Hc2}(a) ). 

\item $ 0 < K < K_{\beta} = 0.018 K_0$.  The helicoid remains the energy minimum at low fields and transforms into the cone phase along line ($\alpha -B$). 

\item $  K_{\beta} < K < K_{A} = 0.050 K_0$. In this narrow interval the cone phase is separated from the alternative modulated states by three first order transitions lines, ($\alpha-A$), ($A-\beta$),
and ($\beta-B$).

\item $  K_{A} < K < K_{B} = 0.363 K_0$. Here the skyrmion lattice area is bounded by the first order lines  ($A -C$) and ($A -\beta -B$) correspondingly into the helicoid and the cone.

\item  $  K_{B} < K < K_{C} = 1.90 K_0$. In this extended interval the evolution of the modulated states follows the scenario characteristic for noncentrosymmetric uniaxial ferromagnets \cite{Bogdanov:1994jmmm}: at low fields (line ($A-C$)) the helicoid transforms by the first-order process into the skyrmion lattice which gradually transforms into a set of isolated skyrmions at critical line $B-C$ (see Fig. \ref{fig:solutions2}).

\item $  K_{C} < K < K_{E} = 2.467 K_0$. In this interval the helicoid directly transforms into the saturated state at line ($C-E$).

\item For $  K > K_{E}$ modulated states are totally suppressed. In this case isolated skyrmions can exist even at zero field in systems with arbitrary large anisotropy.\cite{Bogdanov:1994pss,Kiselev:2011jpd}

\end{enumerate}

\begin{table}[h]
\caption{
\label{table1}
Critical and characteristic points in the $K-H$ phase diagram}
\begin{tabular}{|l|l|l|l|l|l|l|l|l|}

\hline

   & $A$\quad   &  $B$ \quad  &  $C$ \quad &  $E$ \quad &  $\beta$ \quad \;&   $\gamma$ \quad \;&  $\delta$ \quad \; &  $\epsilon$ 
\\

\hline  \hline

 $K/K_0$ &  0.050 & 0.363 & 1.90 & 2.467 & 0.018 & 0.559 & 1.0 & 0.120 \\

\hline 

$H/H_D$ &  0.216 & 0.637 & 0.10 & 0 & 0.360 & 0.441 & 0 & 0 \\
\hline
\end{tabular}
\end{table}

Archetypical $T-H$ phase diagrams for easy-axis systems with  $  K_{B} < K < K_{C} = 1.90$ (case (3)) and for easy-plane system $K \leq 0$ (case (1)) are plotted in Fig. \ref{fig:THtypes} (b), (c), and typical magnetization curves for cases (2), (4), (5) have been calculated in Ref. \onlinecite{Butenko:2010prb}.
\newpage


%
\end{document}